\documentclass[conference]{IEEEtran}
\IEEEoverridecommandlockouts
\usepackage{amsmath,amssymb,amsfonts}
\usepackage{graphicx}
\usepackage{textcomp}

\usepackage{booktabs} % For formal tables
\usepackage{balance}  % to better equalize the last page
\usepackage{graphics} % for EPS, load graphicx instead
\usepackage{mathtools}
\usepackage{algorithm}
\usepackage[noend]{algpseudocode}
\usepackage{url}      % llt: nicely formatted URLs

\usepackage{graphicx} % for EPS use the graphics package instead
\usepackage[font=small,skip=1pt]{caption}
\usepackage{subcaption}
\usepackage{xcolor}
\usepackage{url}
\newcommand{\hide}[1]{}

\usepackage{mdframed}

\renewenvironment{quote}
  {\list{}{\rightmargin=0.35cm \leftmargin=0.35cm}%
   \item\relax}
  {\endlist}

\def\BibTeX{{\rm B\kern-.05em{\sc i\kern-.025em b}\kern-.08em
    T\kern-.1667em\lower.7ex\hbox{E}\kern-.125emX}}
\begin{document}

\title{A Constrained Coupled Matrix-Tensor Factorization for Learning Time-evolving and Emerging Topics
}

\author{\IEEEauthorblockN{Sanaz Bahargam}
\IEEEauthorblockA{\textit{Computer Science Department} \\
\textit{Boston University}\\
%Boston,US \\
bahargam@bu.edu}
\and
\IEEEauthorblockN{Evangelos E. Papalexakis}
\IEEEauthorblockA{\textit{Computer Science Department} \\
\textit{University of California Riverside}\\
epapalex@cs.ucr.edu}
}

\maketitle

\begin{abstract}
%\reminder{TODO for Vagelis: change}
\hide{Despite the rapid increase of new courses offered in online educational platforms, designing a curriculum is still done in traditional ways, manually  in a limited amount of time and based on the prior experience of the instructor. Although there is a massive amount of discussion data  that can  be utilized  to help the instructor design curriculum, little work has been done to take advantage of this data.
%Little work has been done to utilize the massive amount of data available from course discussions or 
In this paper, we introduce a novel designing framework for generating a curriculum from online discussions, which allows us to automatically uncover learning units, identify their sequence and produce a curriculum. Our model allows for efficient and automatic curriculum synthesizing and scales well from small to large discussions forums. Experiments on real data gathered from Physics Stack Exchange and also Programming Stack Exchange forum shows our model is able to discover and characterize topics, track their evolution over a time, and generate meaningful and desirable curriculum. 
}
Topic discovery has witnessed a significant growth as a field of data mining at large. In particular, time-evolving topic discovery, where the evolution of a topic is taken into account has been instrumental in understanding the historical context of an emerging topic in a dynamic corpus. Traditionally, time-evolving topic discovery has focused on this notion of time. However, especially in settings where content is contributed by a community or a crowd, an orthogonal notion of time is the one that pertains to the level of expertise of the content creator: the more experienced the creator, the more advanced the topic. 

In this paper, we propose a novel time-evolving topic discovery method which, in addition to the extracted topics, is able to identify the evolution of that topic over time, as well as the level of difficulty of that topic, as it is inferred by the level of expertise of its main contributors. Our method is based on a novel formulation of  Constrained Coupled Matrix-Tensor Factorization, which adopts constraints well-motivated for, and, as we demonstrate, are essential for high-quality topic discovery.  

We qualitatively evaluate our approach using real data from the Physics and also Programming Stack Exchange forum, and we were able to identify topics of varying levels of difficulty which can be linked to external events, such as the announcement of gravitational waves by the LIGO lab in Physics forum. We provide a quantitative evaluation of our method by conducting a user study where experts were asked to judge the coherence and quality of the extracted topics. 
Finally, our proposed method has implications for automatic curriculum design using the extracted topics, where the notion of the level of difficulty is necessary for the proper modeling of prerequisites and advanced concepts. 

\end{abstract}


%\renewcommand{\Bbb}{\mathbb}
%\algdef{SE}[SUBALG]{Indent}{EndIndent}{}{\algorithmicend\ }%
%\algtext*{Indent}
%\algtext*{EndIndent}

%% Lemma, Theorem, Definiton
\newtheorem{problem}{Problem}
\newtheorem{observation}{Observation}

\newcommand{\squishlist}{\begin{list}{$\bullet$}
  { \setlength{\itemsep}{0pt}
     \setlength{\parsep}{3pt}
     \setlength{\topsep}{3pt}
     \setlength{\partopsep}{0pt}
     \setlength{\leftmargin}{1.5em}
     \setlength{\labelwidth}{1em}
     \setlength{\labelsep}{0.5em} } }
\newcommand{\squishend}{
  \end{list}  }

\newcommand{\codeurl}{\url{www.yyy.com/code.zip}}

\newcommand{\user}{{\ensuremath{\mathbf{u}}}}
\newcommand{\tim}{{\ensuremath{\mathbf{t}}}}
\newcommand{\word}{{\ensuremath{\mathbf{w}}}}

%% problema and algorithm
%\newcommand{\constrainedCMTF}{{ Constrained-CMTF}

%% notations
\newcommand{\vectora}{{\ensuremath{\mathbf{a}}}}
\newcommand{\matrixa}{{\ensuremath{\mathbf{A}}}}
\newcommand{\matrixb}{{\ensuremath{\mathbf{B}}}}
\newcommand{\tensora}{{\ensuremath{\mathcal{A}}}}
\newcommand{\element}{{\ensuremath{a}}}

\newcommand{\tensor}{{\ensuremath{\mathbf{\mathcal{T}}}}}
\newcommand{\core}{{\ensuremath{\mathbf{\mathcal{G}}}}}
\newcommand{\coreSmall}{{\ensuremath{G}}}

\newcommand{\sideMatrix}{{\ensuremath{\mathbf{Y}}}}
\newcommand{\factA}{{\ensuremath{\mathbf{A}}}}
\newcommand{\factB}{{\ensuremath{\mathbf{B}}}}
\newcommand{\factC}{{\ensuremath{\mathbf{C}}}}
\newcommand{\factD}{{\ensuremath{\mathbf{D}}}}
%% size
\newcommand{\sizeModeA}{{\ensuremath{\mathbf{I}}}}
\newcommand{\sizeModeB}{{\ensuremath{\mathbf{J}}}}
\newcommand{\sizeModeC}{{\ensuremath{\mathbf{K}}}}
\newcommand{\sizeSideMatrix}{{\ensuremath{\mathbf{F}}}}

\newcommand{\sizeModeAIter}{{\ensuremath{\mathbf{i}}}}
\newcommand{\sizeModeBIter}{{\ensuremath{\mathbf{j}}}}
\newcommand{\sizeModeCIter}{{\ensuremath{\mathbf{k}}}}
\newcommand{\sizeSideMatrixIter}{{\ensuremath{\mathbf{f}}}}

\newcommand{\sizeFactA}{{\ensuremath{\mathbf{R_1}}}}
\newcommand{\sizeFactB}{{\ensuremath{\mathbf{R_2}}}}
\newcommand{\sizeFactC}{{\ensuremath{\mathbf{R_3}}}}

%% Epsilon Sparsity
\newcommand{\epsSparseA}{{\ensuremath{\epsilon_\factA}}}
\newcommand{\epsSparseB}{{\ensuremath{\epsilon_\factB}}}
\newcommand{\epsSparseC}{{\ensuremath{\epsilon_\factC}}}
\newcommand{\epsSparseD}{{\ensuremath{\epsilon_\factD}}}
\newcommand{\epsSparseCore}{{\ensuremath{\epsilon_\core}}}

%% Epsilon othogonality
\newcommand{\epsOrthogA}{{\ensuremath{\epsilon_\factA}}}
\newcommand{\epsOrthogB}{{\ensuremath{\epsilon_\factB}}}
\newcommand{\epsOrthogC}{{\ensuremath{\epsilon_\factC}}}
\newcommand{\epsOrthogD}{{\ensuremath{\epsilon_\factD}}}

%% datasets  and experiments
\newcommand{\stackex}{{\tt Stack Exchange}}

\newcommand{\rep}{{\ensuremath{\mathbf{rep}}}}

\newcommand{\clusterCenters}{{\ensuremath{\mathbf{C}}}}
\newcommand{\clusterCenter}{{\ensuremath{\mathbf{c}}}}

%% algorithm and problem names
%\newcommand{\paraNS}{\sc PARANS} % greedy algorithm to solve LRemoval problem
%\newcommand{\tukcerNS}{\sc TuckerNS}
%\newcommand{\ourAlgo}{\sc ALSAlgorithm}

\newcommand{\paraNS}{\textsc{\ensuremath{\mathbf{PARAFAC\textendash NS}}}}
\newcommand{\tuckerNS}{ \textsc{\ensuremath{\mathbf{TUCKER3\textendash NS}}}}
\newcommand{\ourAlgo}{ \textsc{\ensuremath{\mathbf{ConCMTF \textendash ALS}}}}

\newcommand{\calX}{{\ensuremath{\mathcal{X }}}}

\newcommand{\argminD}{\arg\!\min}
\newcommand*{\argminl}{\argmin\limits}
\newcommand*{\argmaxl}{\argmax\limits}

%---------------------------------------

%% editing macros
\newcommand{\spara}[1]{\smallskip\noindent{\bf{#1}}}
\newcommand{\mpara}[1]{\medskip\noindent{\bf{#1}}}
\newcommand{\bpara}[1]{\bigskip\noindent{\bf{#1}}}

\section{Introduction}
Traditionally, topic modeling and discovery methods have focused on extracting high quality, interpretable topics that aim to succinctly represent the inherent latent structure within a corpus. Indicatively, there have been different schools of thought on topic extraction, ranging from factorization-based methods \cite{deerwester1990indexing, xu2003document}  to probabilistic graphical models \cite{blei2003latent,steyvers2004probabilistic}.

Recently, there has been significant interest in studying the evolution of topics over time, and this has found particular applications in \cite{wiredmovie} and \cite{he2009detecting}. In general, taking time into account offers the advantage of putting an extracted topic into historical context and can enable the analysis to link the topic to external events that may be related to it.

To the best of our knowledge, the state-of-the-art in time-evolving topic extraction has focused on a notion of ``time'' that pertains to the particular moment that a topic emerged and how it evolved throughout its history within a corpus. However, when we are dealing with topic extraction from community and crowd based platforms, such as \texttt{Stack Exchange}, an additional notion of ``time'' arises. This notion of time is related to the evolution of the user who contributes the content: a relatively new user is more likely to contribute ``entry-level'' content, whereas an experienced user who has already contributed a significant amount of posts, is more likely to create more advanced content. 
%\reminder{do we have a quick way to show that with data??} 
Consider the two following questions posted in \texttt{Stack Exchange} Physics forum. 

\begin{quote}
\textbf{1. Why does centripetal acceleration have a magnitude?}
Assuming that the magnitude of velocity is constant. Why does centripetal acceleration have a magnitude? Since acceleration is the rate of change for velocity and its magnitude remains the same shouldn't we express centripetal acceleration by the angle it changed in the vertical or horizontal over a period of time instead?
\end{quote}

\begin{quote}
\textbf{2. Will the Sun's fast (but subluminal) removal cause gravitational waves?}
 We cannot just remove the sun as it violates energy conservation. We can however let the Sun accelerate fast out of the solar system. Assuming this (unreasonable) scenario, will this fast disappearance of Sun cause any gravitational wave signature? Basically would an experiment such as LIGO be able to measure a gravitational signature of the Sun's removal. 
\end{quote}
The first post is written by a new user who never answered to any others question in the \texttt{Stack Exchange} forum. The question is about ``magnitude of velocity" which is not considered an advanced topic in physics. On the other hand the second post is written by an advanced user who has answer about 70 questions in the forum. The topic of the question is gravitational waves, an advanced topic in physics.
Therefore, information about the ``experience'' of the user who contributed a piece of content in our corpus (measured by the number of post they have already contributed) provides useful information about the level of the content.

Previous work on topic detection has overlooked this notion of time, which relates to user maturity and experience,  and which, as we showcase in this paper, can provide valuable insights on how advanced a particular topic is. In addition to being able to tease out latent concepts of varying levels, these insights are also useful in bootstrapping automated curriculum design approaches such as \cite{agrawal2016toward}  which require a set of concepts to be taught in a curriculum, as well as prerequisite relations for those concepts, which can be given via the user maturity dimension in our topic discovery.

\begin{figure}
 \begin{subfigure}{0.22\textwidth}  
  \centering  
  \includegraphics[width=\linewidth]{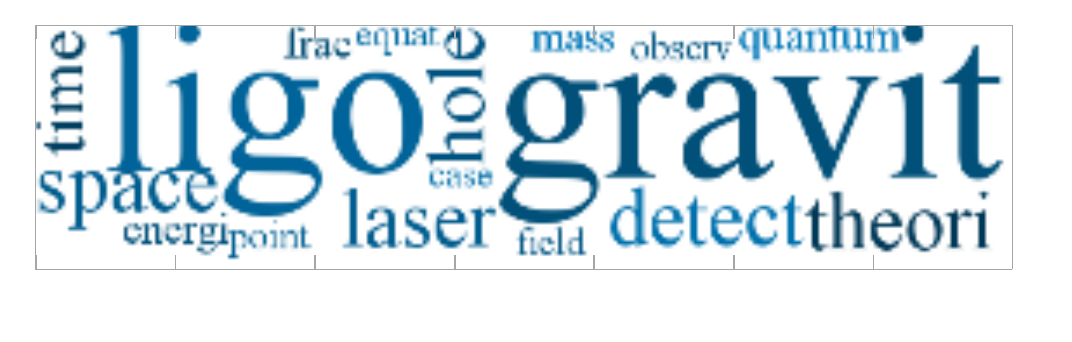}  
    \caption{Words}   
\end{subfigure}   
\begin{subfigure}{0.22\textwidth}  
  \centering  
  \includegraphics[width=\linewidth]{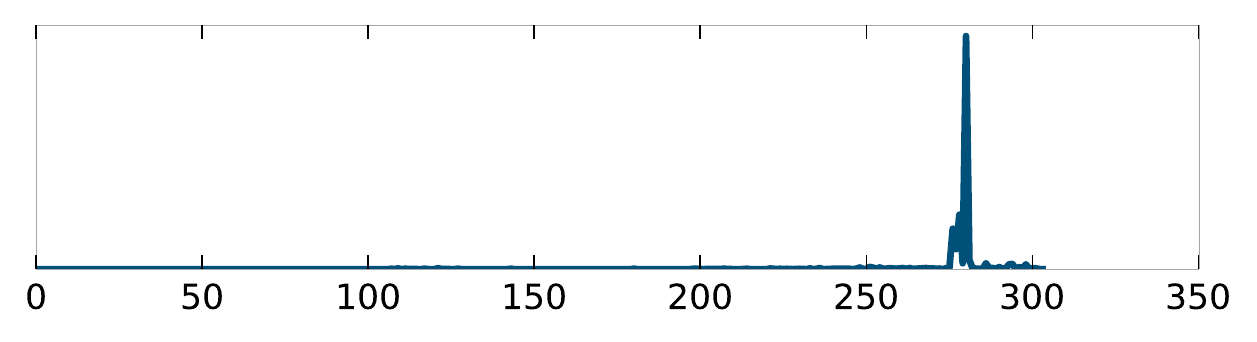}  
    \caption{Time in Weeks}   
\end{subfigure}  

\begin{subfigure}{0.45\textwidth}  
  \centering  
  \includegraphics[width=\linewidth]{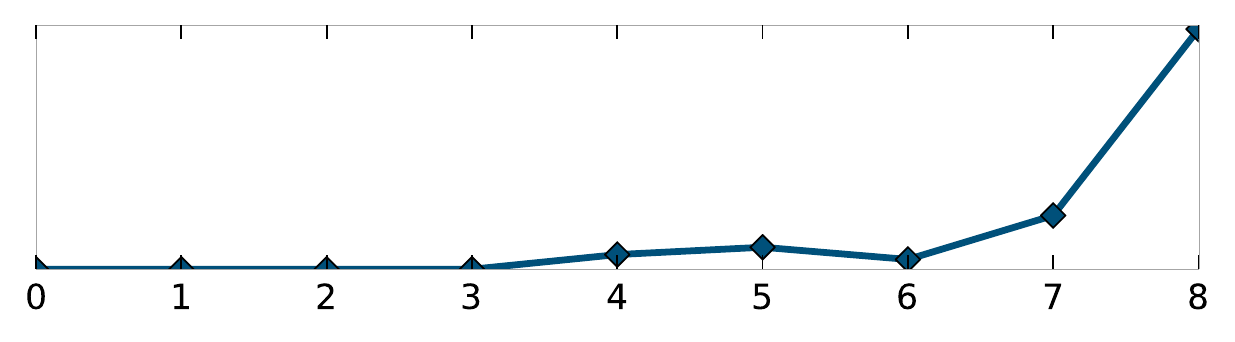}  
    \caption{{\footnotesize``Level of difficulty''}\hide{(measured by Log Post Number, see Section \ref{sec:problem})}}   
\end{subfigure}  
 \caption{\label{fig:ligo}An example of a topic discussed by advances users at a specific time. This pattern indicates topics discussed in response to an external event. The peak of the  ``time'' mode corresponds to February of 2016 when detection of gravitational waves was announced by Ligo lab; furthermore, ``gravitational waves'' is an advanced Physics topic, and our method correctly infers its level of difficulty.}
 \vspace*{-12pt}
\end{figure}

In this paper we introduce an time-evolving topic discovery method, based on Constrained Coupled Matrix-Tensor Factorization, which effectively models time and user maturity/experience towards {\em extracting interpretable topics, their temporal evolution, as well as their level of difficulty}. Figure \ref{fig:ligo} shows a representative such topic detected by our algorithm. The topic corresponds to ``Gravitational Waves Detection by Ligo Lab"; it is clearly an advanced Physics topic, which is indicated by the ``level of difficulty'' aspect of our results, and the topic made its appearance in February of 2016 (as indicated by the ``time'' aspect), which was the date it was announced.

Our contributions are summarized as follows:
\begin{itemize}
\item {\bf Novel Problem Formulation}: We introduce a novel method based on coupled matrix-tensor analysis to discover evolutionary topics and their level of difficulty in online communities. 
%We identify events triggered by external events which are not part of topics evolution and exclude them from the curriculum. We characterize the difficulty of topics and the order that they should appear in the curriculum. 
\item {\bf Constrained Coupled Matrix-Tensor Model}: We propose a novel flexible constrained coupled matrix-tensor factorization model which incorporates sparsity, non-negativity, and orthogonality constraints which are motivated by our topic discovery goal and, as we demonstrate in the experimental evaluation, are essential for the accurate discovery of topics. We derive an efficient Alternating Least Squares algorithm for our proposed factorization model, and in order to promote reproducibility and further research, we release our code at \url{https://github.com/ConCMTF/ConCMTF}.
\item {\bf Evaluation on Real Data}: We qualitatively evaluate our method in comparison to the baseline approaches on real and public data from the Physics and also Programming forum of Stack Exchange. In particular, we demonstrate the power of the proposed method in discovering easy-to-interpret time-evolving topics and their level of difficulty. 
\item {\bf User Study}: For the quantitative analysis of our method, we conduct a user study among 10 domain experts in Physics who judged the quality, interpretability, and coherence of our results.
%\item {\bf Other Applications}: Although this paper focuses on the application of our method for topic extraction, our proposed technique has proven effective on other datasets and other applications as well. Due to lack of space, we omitted the experiments on other datasets. Interested readers can find more about experiments on other datasets in  \url{http://cs-people.bu.edu/bahargam/cmtf/}.
\end{itemize}

%\reminder{It's good if we have a cool picture of our most impressive results here in the first page - usually the reviewer forms an opinion about the paper after reading the intro, so we should impress them right away! :)}

\spara{Roadmap}
The paper is structured as follows. We start with a discussion of related works in Section \ref{prelim}. In Section \ref{prelim},  we provide certain our notations and background on tensor factorization. 
We formally define our framework in Section \ref{problem}. Our algorithm for the problem is described
in Section \ref{algo}. 
Section \ref{results} presents the results and  empirical evaluations.
Finally, we conclude with a summary and directions for future work in Section \ref{conclusion}. A version of this paper appears in the Proceedings of the 2018 IEEE/ACM International Conference on Advances in Social Networks Analysis and Mining \cite{Bahargam:2018}.

\section{Related Work} \label{related}
\subsection{Tensor Decomposition}
Our problem is clearly related to a large body of work both in the domain of
tensor decomposition as well as topic extraction. We give an overview of the existing
work in each domain below.

\textbf{Tensor Decomposition:}
The tensor decomposition technique was first introduced by Hitchcock \cite{hitchcock1927expression} in 1927.
Many variations of the tensor decomposition problem have been studied; among them, there are two particular tensor decompositions that can be considered as higher-order extensions
of the matrix singular value decomposition: (1) CANDECOMP/PARAFAC (or CP for short), and (2) the Tucker decomposition. For an overview of tensor decomposition techniques, see Kolda et al.   \cite{kolda2009tensor},
and Papalexakis et al. \cite{papalexakis2016tensors}. 
%For the applications of Tensor decomposition in exploratary data analysis see Agrawal et al. 

\hide{
One of the most common and widely used family of algorithms for tensor decompositions (and specially CP decomposition)
is the Alternating Least Squares (ALS) algorithms. ALS algorithms find a decomposition by solving for each factor at
a time while fixing all the other factors. In addition to the ALS algorithms, all-at-once algorithms can be used.
Acar et al. \cite{acar2011all} apply first-order gradient optimization methods such as limited-memory BFGS.
Phan et al. \cite{phan2013low} exploit the structure of the Hessian for an efficient second-order
optimization using Newton's method. A study close to ours is the work by Gilpin et al. \cite{gilpin2016some} that
studied the role discovery problem. Their framework models role detection as a constrained non-negative matrix
factorization problem, where the guidance is provided as convex constraints and specified per role. 
}

Another tensor decomposition technique is Coupled Tensor Factorization which was introduced by Smilde et al.  \cite{smilde2000multiway} in the area of Chemometrics.
Since then there has been significant development of such coupled models, either when matrices are coupled
together \cite{singh2008relational} or when matrices and tensors are coupled  \cite{ermics2015link,narita2011tensor}. A notable example of using
the coupled tensor factorization is a recent work by Papalexakis et al. \cite{papalexakis2014turbo} which seeks
to identify coherent regions of the brain using a (noun, voxel, person) tensor and a (noun, feature) matrix.

\textbf{Topic Extraction:}
The topic extraction problem has been rigorously studied in the past. Among the existing methods for solving
this problem, is the family of Markov chain-based topic-extraction methods
\cite{blei2006dynamic,wei2007dynamic,zhang2010evolutionary,ren2008dynamic,ahmed2012timeline}.
In \cite{blei2006dynamic} a Dynamic Topic Modeling tool is proposed which captures the topic evolution
in a collection of documents that is organized sequentially into several discrete time periods, and then
within each period an LDA model is trained on its documents. The Gaussian distributions are used to tie a collection of LDAs by chaining the Dirichlet prior and the natural parameters of each topic.

Mei et al. \cite{Mei:2005:DET} suggested a mixture model to extract the subtopics in weblog collections, to identify and 
track topics in time-stamped text data. In a similar work by Morinaga et al. \cite{Morinaga:2004:TDT} finite mixture
model is used to represent documents at each discrete time interval. In their model, a topic changes on certain documents
if the topic mixtures drift significantly from the previous ones. Kandylas et al. \cite{kandylas2008finding} analyzed the
evolution of knowledge communities using a method called Streemer which focuses time-evolving clusters.

In a work by Aggarwal et al. \cite{aggarwal2006framework} on topic modelling of data streams, a fixed number of
$k$ clusters (topics) are maintained over time. If a new document arrives that is far from all existing clusters, it can
become a new cluster. Liu et al. \cite{liu2008clustering} take a similar approach except instead of using single words as
document features, they use multiword phrases as topic signatures. A drawback of these methods is they consider a-priori
fixed feature space per topic. 

Most works on extracting time-evolving topics, use post-discretized or pre-discretized time analysis. Post-discretized
methods fit a topic modeling to documents without considering time and then documents are sorted in time by slicing them
into discrete subsets \cite{griffiths2004finding}. However, in pre-discretized methods \cite{wang2005group, song2005modeling},
the documents are first sliced into discrete time slices, and then a topic model is fitted to each slice separately. 

Our work is different from previous topic evolution methods as the majority of previous attempts have considered and defined
a time span for each topic (e.g. ~\cite{Mei:2005:DET} ). %  which determines the starting and termination time stamps of the topic. 
In these methods, the extracted topic is highly dependent on how the time-spans are defined. In our work, we do not
need to define a time-span. 
%In addition, in a vast body of work, the whole dataset is partitioned into sub-collections with fixed or variable time intervals. In our case we cannot partition question/answers into sub-collections since when a thread starts (a question is posted) it may continue for days or even months. 
Another shortcoming of the vast majority of existing works is that a-priori fixed feature space is considered for each topic, whereas in
our model we define a topic as a collection of words that appear together. Another advantage of our work is beside
finding time-evolving topics, our method is capable of detecting bursty or conventional topics, such as ``Japan tsunami''
or ``democratic convention'' since we consider both the time and post number as modes in our data.

\section{Preliminary Definitions} \label{prelim}
In this section, we provide the notation we will use throughout the paper. In addition, we explain two popular tensor decompositions  canonical or PARAFAC  and Tucker 3. 
%\subsection{Notations}

\textbf{Notations}:
Vectors are denoted by boldface lower case letters, e.g \vectora. Matrices are denoted by boldface Capital letters, e.g \matrixa. Tensors are denoted by Calligraphic letters, e.g \tensora.  

%An entry of a vector \vectora, a matrix \matrixa, or a tensor \tensora\ is denoted by $\element_{i}, \element_{ij}, \element_{ijk}$.  Let  $\tensor_{\matrixa}$ be the matricization of \tensor in the first mode. The Kronecker product of two matrices is denoted by $\matrixa \otimes \matrixb$. The $n$ mode product is denoted by $\times_{n}$. The outer product of two vectors is denoted by $\circ$.  $\parallel \matrixa \parallel_{F}$ denotes the Frobenius norm of matrix \matrixa.  Moore-Penrose Pseudoinverse of \matrixa is denoted by $\matrixa^{\dagger}$

\begin{center}
\footnotesize
 \begin{tabular}{||c | c ||} 
 \hline
 \textbf{Symbol} & \textbf{Definition} \\ [0.5ex] 
 \hline\hline
  $\element_{i}$ & An entry of a vector \vectora (same for matrix and tensor)\\  \hline
 % $\element_{ij}$ & An entry of a matrix \matrixa \\  \hline
 % $\element_{ijk}$ & An entry of a tensor \tensora \\  \hline
   $\tensor_{\matrixa}$ & Matricization of \tensor in the first mode \\ \hline
   $\matrixa \otimes \matrixb$ &  Kronecker product of two matrices  \\ \hline
   $\times_{n}$ & The $n$ mode product \\ \hline
   $\circ$ & Outer product of two vectors  \\ \hline
   $\parallel \matrixa \parallel_{F}$  & Frobenius norm of matrix \matrixa \\ \hline
    $\matrixa^{\dagger}$ & Moore-Penrose Pseudoinverse of \matrixa \\ [1ex]  \hline
\end{tabular}
\end{center}

%\subsection{Tensor Decomposition}

%\subsection{CP/PARAFAC Tensor Decomposition}
\textbf{CP/PARAFAC Tensor Decomposition:}
Given tensor \tensor, we can analyze into a sum of $F$ rank-one factors  using the CP/PARAFAC decomposition.
\hide{
and obtain a sum of $F$ rank-one tensors as the following:
$$
\footnotesize
\tensor \approx \sum \limits_{f=1}^F \mathbf{a}_f  \circ \mathbf{b}_f \circ \mathbf{c}_f
$$
}
Typically, in order to compute the CP/PARAFAC decomposition, we solve the following optimization problem:
\[ \footnotesize
\min\limits_{a, b, c}  \parallel \tensor - \sum\limits_{f}  \mathbf{a}_f \circ \mathbf{b}_f \circ \mathbf{c}_f   \parallel_{F}^{2}
\]
which minimizes the Frobenius norm of the difference between the tensor and the model, where $\mathbf{a}_f$, $\mathbf{b}_f$, and $\mathbf{c}_f$ are latent vectors that correspond to words, time, and post numbers.  

\hide{
Figure \ref{fig:parafac} illustrates how we model and analyze the forum data using PARAFAC decomposition. 

\begin{figure}[!ht]
	\begin{center}
		\includegraphics[width=0.15\textwidth]{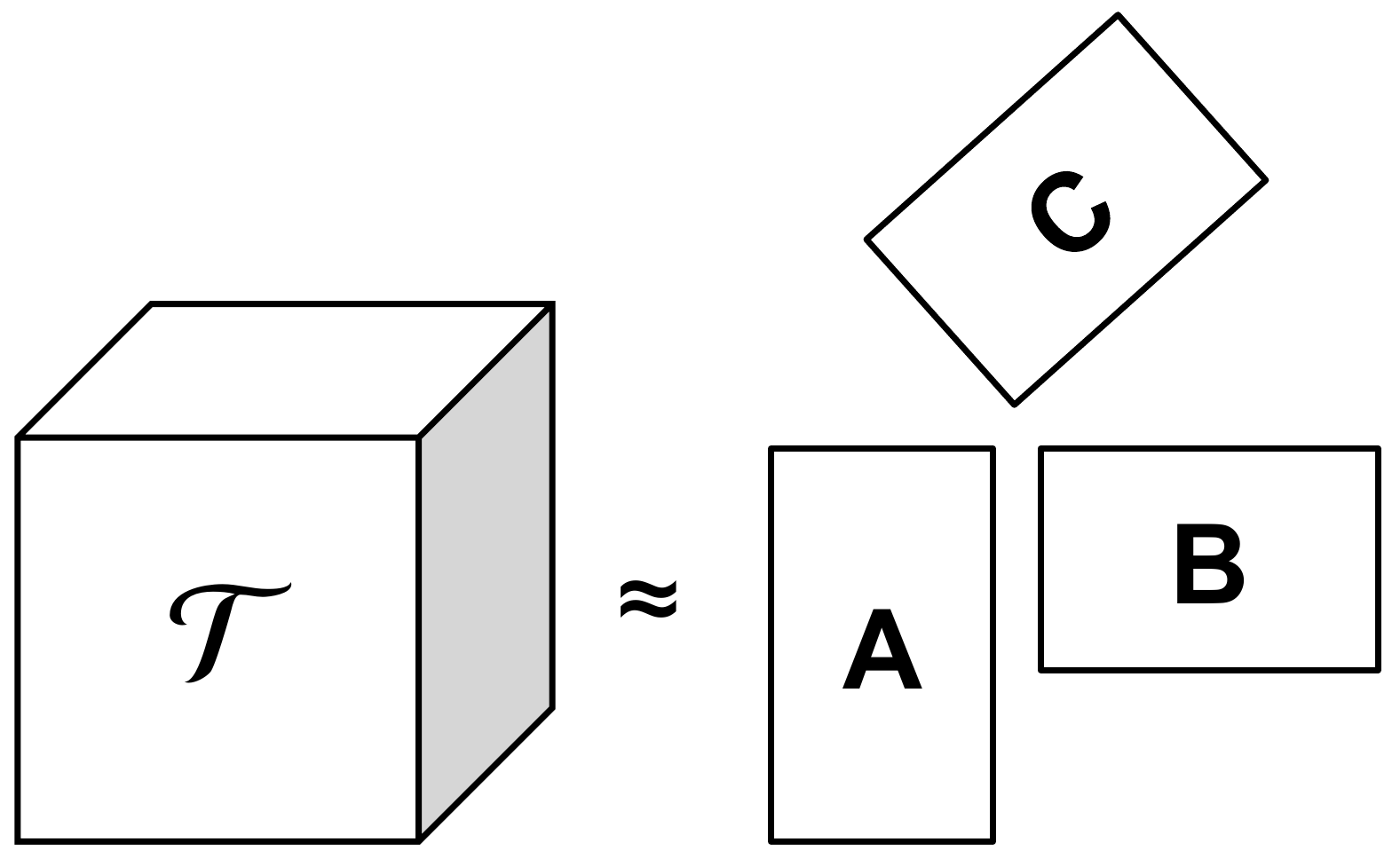}
		\caption{The PARAFAC decomposition}
		\label{fig:parafac}
	\end{center}
\end{figure}
}

To decide about the right number of latent factors ($F$), we used AutoTen \cite{papalexakis2016automatic} which allows us to find more structured latent embeddings in the data. 
Each latent factor of the embeddings defines a pattern in the forum meaning a set of words (topic) used in a specific time in specific post numbers. If a set of words only appear in a specific short time period (with a bursty distribution), we consider it a topic triggered by an external event at a specific time, Figure \ref{fig:ligo}. On the other hand, if a set of words appear all the time but only specific post numbers we consider it as a part of the evolutionary topics. The post number of these topics determines the {\em difficulty of the topic}. For example, if a topic appears in low post numbers, it corresponds to basic topics only discussed by newbies, Figure \ref{fig:coup_parafac_c1word}.  Whereas  Figure \ref{fig:coup_parafac_c3word} shows that the topics mainly appearing in  high post numbers written by advanced users who have contributed a lot in the past.
PARAFAC is unique under mild conditions.  This is important because it allows us to uniquely unravel a
large number of possibly overlapping co-clusters that are hidden in the data. As a result, having a tensor with (word, time, post number) modes, one can view  PARAFAC as a soft clustering that detects  groups of words that tend to appear together in certain time intervals and specific post numbers. Thus, in this case co-clustering  is taking advantage of the ternary relationship between words, time stamps, and post numbers which makes it a good candidate for topic modelling applications. 
In a similar work, Agrawal et al. \cite{Agrawal:2017:HWS, Agrawal:2015:OGB:2817946.2820604, Agrawal:2015:SDW:2740908.2743060, Agrawal:2015:WSN:2783258.2788571}  used PARAFAC as a co-clustering to model the comparison between the results of different search engines, based on emerging latent topics.  For a set of queries, they create a (query, keyword, date, search engine) tensor and use the CP decomposition to create latent representations of search engines in the same space.

%\subsection{Tucker 3 Tensor Decomposition}
\textbf{Tucker 3 Tensor Decomposition:}
In addition to CP/PARAFAC, the other most widely used tensor decomposition model is the Tucker model \cite{tucker1966some}. In the original paper, Tucker introduced three models; in this paper we are going to focus on the third one, also known as Tucker 3, which can be seen as a generalization of CP/PARAFAC. In Tucker 3 the tensor is decomposed into
$ \footnotesize
\tensor \approx \sum\limits_{i} \sum\limits_{j} \sum\limits_{k}  \coreSmall_{ijk} * \mathbf{a}_i \circ \mathbf{b}_j  \circ \mathbf{c}_k   
$
 where now the factor vectors are combined using a core tensor $\core$. The $(i,j,k)$ entry $\coreSmall_{ijk}$ of the core tensor is indicating the interaction between the $(i,j,k)$ latent factors. We can write CP/PARAFAC as a special Tucker 3 model where th he core is super-diagonal, i.e., it only has non-zero values in $(i,i,i)$.

The existence of the core tensor $\core$ in Tucker 3 is key to our application. Due to this core, Tucker 3 is able to capture {\em interactions between latent components} which, as we will see in the rest of the paper, are important for topic discovery. Sparsity on core: Imposing sparsity constraint on core tensor, as in e.g., \cite{gilpin2016some}, improves the interpretability of the components since fewer interactions are included. 
%Non-sparse core tensor makes the interaction and relationship between factors ambiguous specially in decision based applications such as classification.  A sparse core tensor with fewer interactions reduces this ambiguity.  In addition it yields a more compact representation and compresses the data by considering only non-zero values of the core tensor. 

In tensor/matrix form, the Tucker 3 model can be written as
$
\tensor \approx \core_{\times_{3}} \factC_{\times_{3}} \factB_{\times_{2}} \factA_{\times_{1}}
$
where ${\times_{N}}$ is the $N$-mode tensor-matrix product.
% and essentially multiplies the $N$-th mode of a tensor with the corresponding mode of a matrix, similar to matrix multiplication. Tucker 3 is shown pictorially in Figure \ref{fig:tucker}.
\hide{
\begin{figure}[!ht]
	\begin{center}
		\includegraphics[width=0.15\textwidth]{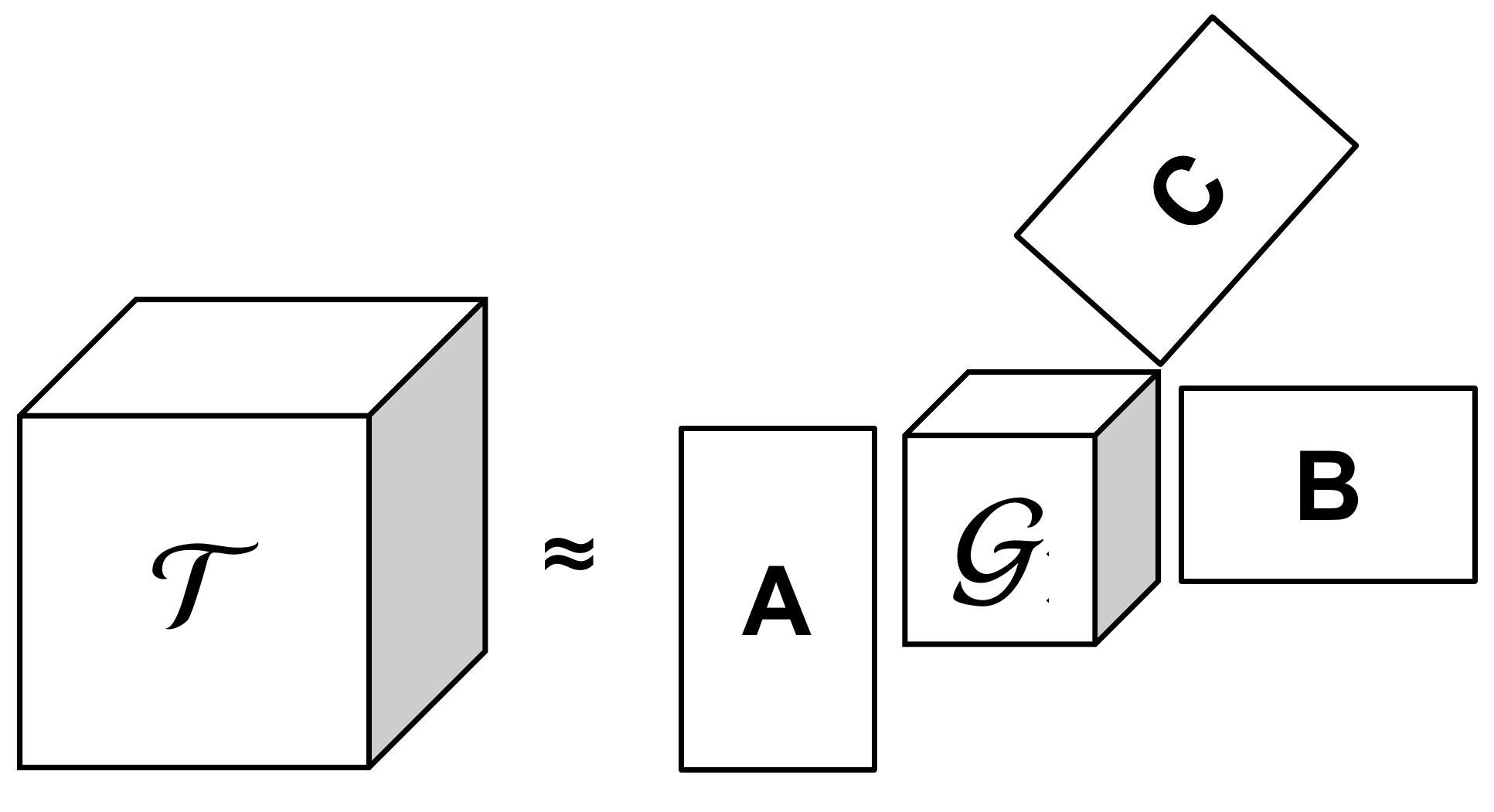}
		\caption{The Tucker 3 decomposition}
		\label{fig:tucker}
	\end{center}
\end{figure}
}

\section{Proposed Constrained Coupled Matrix-Tensor Factorization} \label{problem}
\label{sec:problem}
In a wide variety of applications, we have data that form a tensor
and have side information or metadata that may form matrices or other tensors.
For instance, suppose we have a (word, time, post number) tensor that indicates how many times a word was used in a specific time and specific post numbers. Usually, question answering platforms also have some metadata on the
questions/answers, for instance tags of the questions, that can form a (words, tags) matrix. Thus  we have a third-order tensor, $\tensor \in {\mathbb R}^{\sizeModeA \times \sizeModeB \times \sizeModeC}$ and a matrix $\sideMatrix \in {\mathbb R}^{\sizeModeA \times \sizeSideMatrix}$, coupled in the first mode of each and there is a one-to-one correspondence of elements in the first mode of the tensor and the matrix (``word'' mode in our case). The coupled-matrix and tensor factorization (CMTF) algorithms  jointly factorizes multiple data sets in the form of higher-order tensors and matrices by extracting a common latent structure from the shared mode.
Imposing a Tucker model yields:

\hide{
\begin{equation}\label{eq:coupled}
\min\limits_{a_{\sizeModeAIter}, b_{\sizeModeBIter}, c_{\sizeModeCIter}, d_{\sizeModeAIter}}  \parallel \tensor - \sum\limits_{i} \sum\limits_{j} \sum\limits_{k}  \coreSmall_{ijk} * a_i \circ b_j  \circ c_k   \parallel_{F}^{2}  + \parallel \sideMatrix - \sum\limits_{i} a_i d_{i}^{T}  \parallel_{F}^{2}
\end{equation}

More concisely
}

\begin{equation}\label{eq:coupledConc}
\min  \parallel\tensor - \core_{\times_{3}} \factC_{\times_{3}} \factB_{\times_{2}} \factA_{\times_{1}}\parallel_{F}^{2} + \parallel \sideMatrix - \factA\factD^{T}  \parallel_{F}^{2} 
\end{equation}

On the other hand if we impose a PARAFAC decomposition, we have
\begin{equation}\label{eq:coupled_cp}
\min\limits_{a_{r}, b_{r}, c_{r}, d_{r}}  \parallel \tensor - \sum\limits_{r}   a_r \circ b_r  \circ c_r   \parallel_{F}^{2}  + \parallel \sideMatrix - \sum\limits_{r} a_r d_{r}^{T}  \parallel_{F}^{2}
\end{equation}

\hide{
\begin{figure}[!ht]
	\begin{center}
		\includegraphics[width=0.15\textwidth]{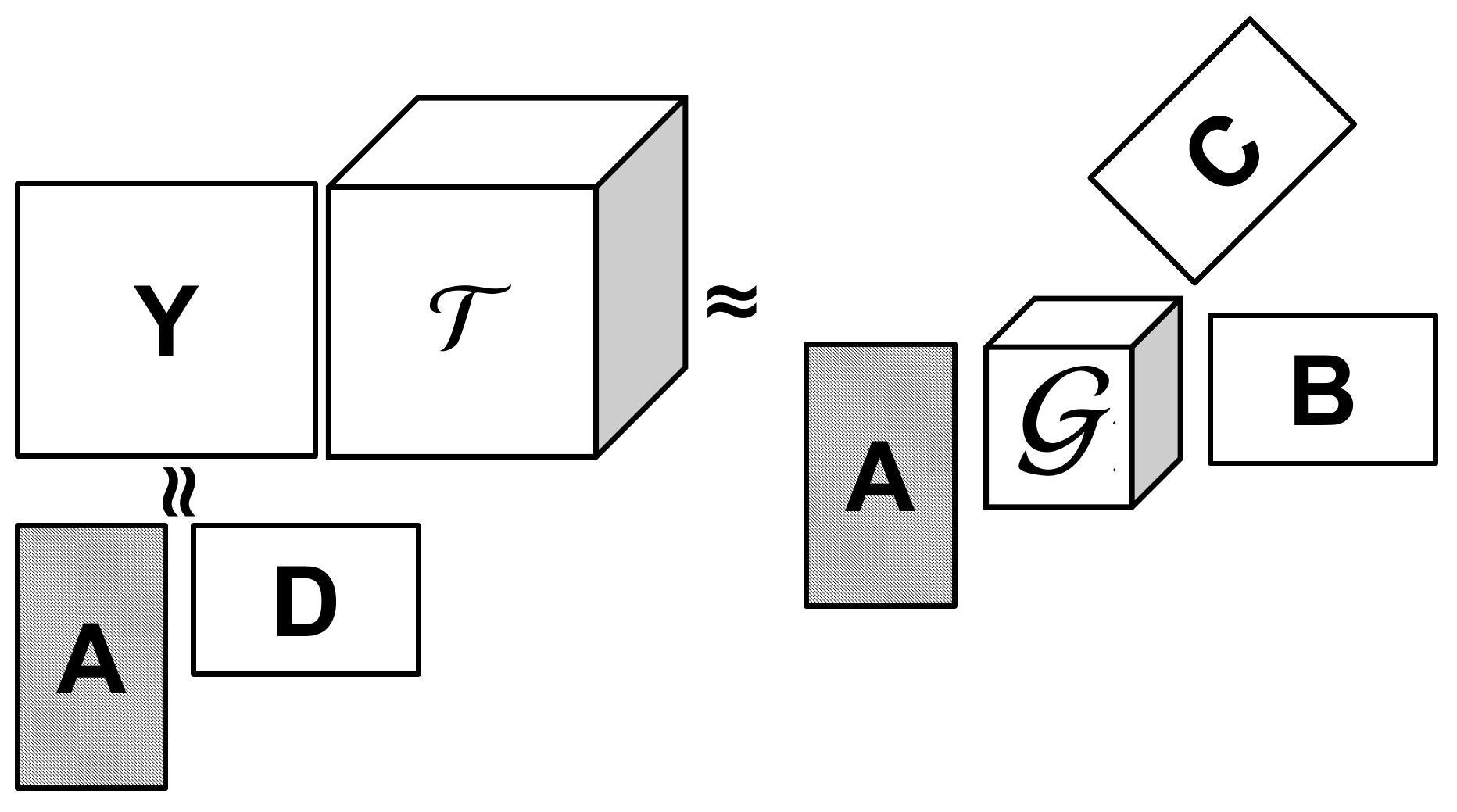}
		\caption{Coupled Matrix-Tensor Factorization with a Tucker 3 model for the tensor}
		\label{fig:coupled}
	\end{center}
\end{figure}
}
The existing work on coupled-matrix tensor factorization only considers non-negativity constraints, e.g. $\factA \geq 0$. 
Non-negativity is an important feature of latent factors since many real world tensors have non-negative values and hidden components have a physical meaning only when non-negative. 

Although non-negativity improves interpretability,  in many applications it is not enough to make sense of the data. When the goal of factorization is to find the latent topics within the tensor and the matrix, we would like to find as many non-overlapping structures as possible. Non-overlapping latent components directly imply that the latent topics are concise and hence interpretable.  We can control the amount of overlap in latent components by imposing orthogonality constraint on each latent component. This means for the first mode, we would like the columns of the latent component \factA\ to be orthogonal, $\forall i, j \quad \factA_{i}^{T} \factA_{j} \leq \epsOrthogA \quad i \neq j$. If $ \epsOrthogA$ is set to $0$, this implies latent components should be completely orthogonal, while values greater than $0$ means some overlap is allowed. 

Furthermore, in practice we desire the factors to be sparse as well. Sparsity constraints improve parsimony and offer a simpler and hence more interpretable model.  We can impose sparsity constraint on all factors and on the core tensor as well. We enforce the sparsity constraint by imposing constraint on $\ell_{1}$ norm of each column in factor matrices and on the core tensor. Enforcing sparsity on each column of the factor matrices means sparsity is imposed uniformly on each latent component for each mode. 
Sparsity becomes specially  favorable  when it is imposed on the core tensor; meaning only a few latent components interact with each other. This removes redundancy and achieve compact sparse representations of the core and hence the core tensor will be easily interpretable. 

To the best of our knowledge we are the first to introduce the constraint coupled-matrix tensor factorization problem with non-negativity, sparsity and orthogonality constraints.  
Our intuition and constraints  are captured in a formal definition as follows.
\begin{problem}[]\label{prob:constrainedCMTF}
Given a tensor $\tensor \in {\mathbb R}^{\sizeModeA \times \sizeModeB \times \sizeModeC}$, auxiliary matrix $\sideMatrix \in {\mathbb R}^{\sizeModeA \times \sizeSideMatrix}$, and number of factors for each component \sizeFactA, \sizeFactB, \sizeFactC,  find the components \factA, \factB, \factC, \factD, and tensor \core\ such that
$$\min  \parallel\tensor - \core_{\times_{3}} \factC_{\times_{3}} \factB_{\times_{2}} \factA_{\times_{1}}\parallel_{F}^{2} + \parallel \sideMatrix - \factA\factD^{T}  \parallel_{F}^{2},$$
$$  \text{Subject to: }  \text{For each factor  } F \in \lbrace \factA, \factB, \factC, \factD \rbrace$$
$$ F \geq 0 \text{ and } \forall i \parallel F_{i} \parallel_{1} \leq \epsilon_{F1}, 
\text{ and } \forall i, j \quad F_{i}^{T} F_{j} \leq \epsilon_{F2 } \quad i \neq j  $$
$$ \text{For core tensor } \core, \core \geq 0,  \parallel \core \parallel_{1} \leq \epsilon_{\core},$$
\end{problem}

%In our problem we do not consider orthogonality for core tensor, since we want to allow interaction between same factors, but only a few of them. 
For the sake of interpretation, it is enough for core to be sparse, having a few non-zero elements. Lifting orthogonality constraint from core means we allow interaction between same factors, but only a few factors to interact with each other. 
%However with some changes, we can  imposed orthogonality constraints on core tensor in the same way we imposed orthogonality constraints on components. 
%In our experiments,  we only consider non-negativity and orthogonality for factors and non-negativity and sparsity for core tensor. 

\section{Proposed Algorithm} \label{algo}
Even though, tensor decomposition is an NP-hard problem, here we provide an Alternating Least Squares (ALS) algorithm which solves the Constrained Coupled-Matrix Tensor factorization problem and converges to a locally optimal solution (as a property of the family of ALS algorithms).
Alternating Least Squares  is probably the most widely used algorithm and dates
back to the original papers by Carroll et.al \cite{carroll1970analysis} and Harshman \cite{harshman1970foundations}.
Alternating Least Squares method has been shown to be very efficient and competitive for PARAFAC decomposition \cite{tomasi2006comparison} and in practice works well \cite{kolda2009tensor}. 
This approach has the advantage that it can be applied to large scale problems.  Using ALS method,  we solve for each factor at a time while fixing all other factors.  If we seek to estimate $\factA$, it turns out that we need to concatenate the two pieces of the data \tensor and \sideMatrix, whose rows refer to matrix \factA, that is the matricized tensor $\tensor_{\factA}$ and matrix \sideMatrix, and we can then solve for \factA\ as

\begin{equation}\label{eq:coupled2}
\factA 
=
\begin{bmatrix} \tensor_{\factA}  \\
\ \sideMatrix
  \end{bmatrix}^{T}
  \Bigg(
  \begin{bmatrix} \core_{\factA}(\factB \otimes \factC) \\
  \factD
  \end{bmatrix}^{\dagger}
  \Bigg)^{T}
\end{equation}

Algorithm \ref{algo_als} shows our ALS algorithm to solve the constrained coupled-matrix tensor factorization, \ourAlgo.  
These constraints include non-negativity, sparsity and orthogonality imposed by $\factA \geq 0$, $\forall i \parallel \factA_{i} \parallel_{1} \leq \epsSparseA$ and  $\forall i, j \quad \factA_{i}^{T} \factA_{j} \leq \epsOrthogA \quad i \neq j$  respectively. 
Rather than alternating to solve each factor completely, we solve for each column of each factor independently. This is possible since the columns of each factor are independent and the constraints we consider can be specified to each column as well. A column in the factor of first mode, \factA, indicates a group of words and a column in \factB indicates a specific weeks in the lifetime of the forum.  It is important to note the effect of  specifying sparsity constraints on the columns rather than the whole matrix.  This means {\em sparsity will be spread uniformly across the whole matrix}.  
It is worth mentioning that our algorithm can allow any convex constraints to be placed for each factor. 
Another advantage of our algorithm is that it can be easily used for  PARAFAC decompositions instead of Tucker3 with minimal changes. To achieve this, instead of initializing core to random values in Line \ref{algo_als:init}, we set the core tensor to a super diagonal tensor. In addition, there is no need to estimate core tensor in each iteration and hence Line \ref{algo_als:solveCore1} and \ref{algo_als:solveCore} can be removed from the algorithm. 

\begin{algorithm}[ht!]
%\footnotesize 
\footnotesize
\begin{algorithmic}[1] 
\Statex {\bf Input:}  The tensor $\tensor \in {\mathbb R}^{\sizeModeA \times \sizeModeB \times \sizeModeC}$ and auxiliary matrix $\sideMatrix \in {\mathbb R}^{\sizeModeA \times \sizeSideMatrix}$
\Statex {\bf Output:}  Coupled Decompositions $\factA \in {\mathbb R}^{\sizeModeA \times \sizeFactA}, \factB \in {\mathbb R}^{\sizeModeB \times \sizeFactB}, \factC \in {\mathbb R}^{\sizeModeC \times \sizeFactC}, \factD \in {\mathbb R}^{ \sizeSideMatrix \times \sizeFactA  }$
\State Initialize $\factA, \factB, \factC, \factD, \core$ to non-negative random values \label{algo_als:init}
\While{convergence criterion is not met} \label{algo_als:p2_start}
\State $\factA \leftarrow \argminD\limits_{\factA} || [\tensor_{\factA} \quad \sideMatrix] - \factA [\core_{\factA}(\factC \otimes \factB)^{T} \quad \factD^{T} ] ||_{Fro}$\\ \indent\indent\indent Subject to: $\factA \geq 0$ and $\forall i \parallel \factA_{i} \parallel_{1} \leq \epsSparseA$ \\ \indent\indent\indent and $\forall i, j \quad \factA_{i}^{T} \factA_{j} \leq \epsOrthogA \quad i \neq j$ \label{algo_als:contsraints}
\State Normalize the columns of $\factA$

\State $\factB \leftarrow \argminD\limits_{\factB} ||\tensor_{\factB} - \factB \core_{\factB} (\factC \otimes \factA)^{T} ||_{Fro}$\\ \indent\indent\indent Subject to: $\factB \geq 0$ and $\forall i \parallel \factB_{i} \parallel_{1} \leq \epsSparseB$ \\ \indent\indent\indent and $\forall i, j \quad \factB_{i}^{T} \factB_{j} \leq \epsOrthogB \quad i \neq j$
\State Normalize the columns of $\factB$

\State $\factC \leftarrow \argminD\limits_{\factC} ||\tensor_{\factC} - \factC \core_{\factC} (\factB \otimes \factA)^{T} ||_{Fro}$\\ \indent\indent\indent Subject to: $\factC \geq 0$ and $\forall i \parallel \factC_{i} \parallel_{1} \leq \epsSparseC$ \\ \indent\indent\indent and $\forall i, j \quad \factC_{i}^{T} \factC_{j} \leq \epsOrthogC \quad i \neq j$
\State Normalize the columns of $\factC$

\State $\factD \leftarrow \argminD\limits_{\factD} ||\sideMatrix - \factA \factD^{T} ||_{Fro}$\\ \indent\indent\indent Subject to: $\factD \geq 0$ and $\forall i \parallel \factD_{i} \parallel_{1} \leq \epsSparseC$ \\ \indent\indent\indent and $\forall i, j \quad \factD_{i}^{T} \factD_{j} \leq \epsOrthogC \quad i \neq j$
\State Normalize the columns of $\factD$

\State $\core \leftarrow \argminD\limits_{\core} ||vec(\tensor) - (\factC \otimes \factB \otimes \factA)vec(\core)||_{Fro}$ \label{algo_als:solveCore1}  \\ \indent\indent\indent Subject to: $\core \geq 0$  and $\parallel \core \parallel_{1} \leq \epsSparseCore$ \label{algo_als:solveCore}

\EndWhile \label{algo_als:p2_end}

\State return \factA, \factB, \factC, \factD, \core
\end{algorithmic}
\caption{\label{algo_als} The Alternating Least Squares for Constrained Coupled Matrix-Tensor Factorization \ourAlgo}
\end{algorithm} 
\subsection{Running Time:}
Each step of our algorithm can be solved by any convex or least squares (LS) solvers; 
If we chose a LS solver or a non-negative least squares (NNLS) solver such as the one in the N-way Matlab Toolbox \cite{andersson2000n}, we would subsequently need to transform the unconstrained NNLS solution into a constrained one by using projected gradient descent. However, 
For flexibility, we used CVX,  a convex solver. 
In practice we observed that CVX is faster than using Least Squares solvers and projected gradient decent method combined. 
Each iteration of our algorithm has linear complexity with respect to the number of factors, and number of modes.
%and in practice it takes less than a minute to compute the factors. 
In addition, our algorithm solves for each column of each matrix independently and hence makes our algorithm faster. 
More precisely, assuming a Parafac decomposition on a tensor of size $[n,n,n$] with m factors, the running of to solve for each factor at each iteration will be $O(m * n ^3)$ in which $m$ is the number of factors and $n$ is the number of variables in each column.  In theory, the convex  programming can be solved in the cubic number of variables ($O(n^3)$) \cite{ye1989extension} but in practice it runs faster. So the total running time of our algorithm is $iterations * O(m * n^3)$.
In addition our algorithm converges in less than a few tens of iterations. Thus, the whole running time is reasonable.

\section{Results} \label{results}
We focus on question and answers related to the field of physics and python programming in \texttt{Stack
Exchange}. 
 We start by discussing our datasets in details, and then we present how we apply our tensor decomposition techniques on this datasets to find topics.

\subsection{Data}
\begin{figure*}
\captionsetup[subfigure]{aboveskip=-1pt,belowskip=-1pt}
  \centering
    \begin{subfigure}[b]{0.31\textwidth}
                \includegraphics[width=\linewidth]{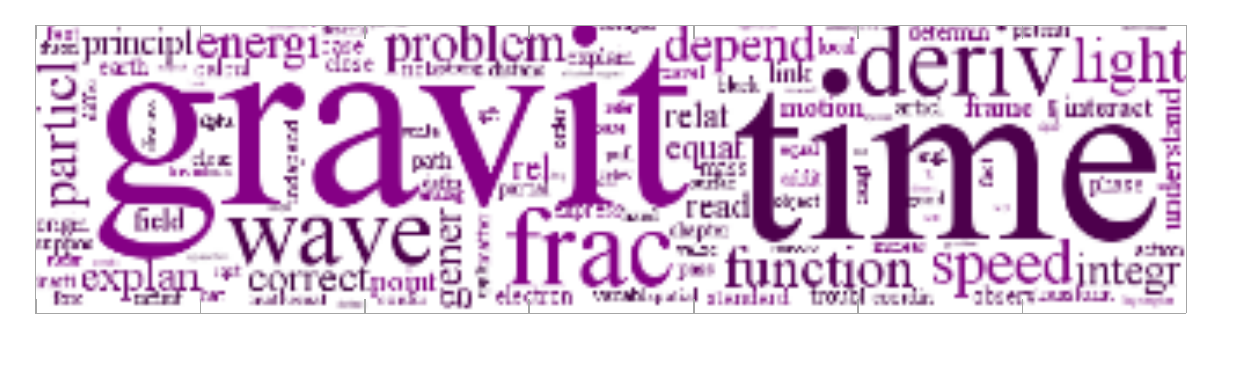}
                \caption{Words}
                \label{fig:parafac_c1word}
        \end{subfigure}%
        \begin{subfigure}[b]{0.31\textwidth}
                \includegraphics[width=\linewidth]{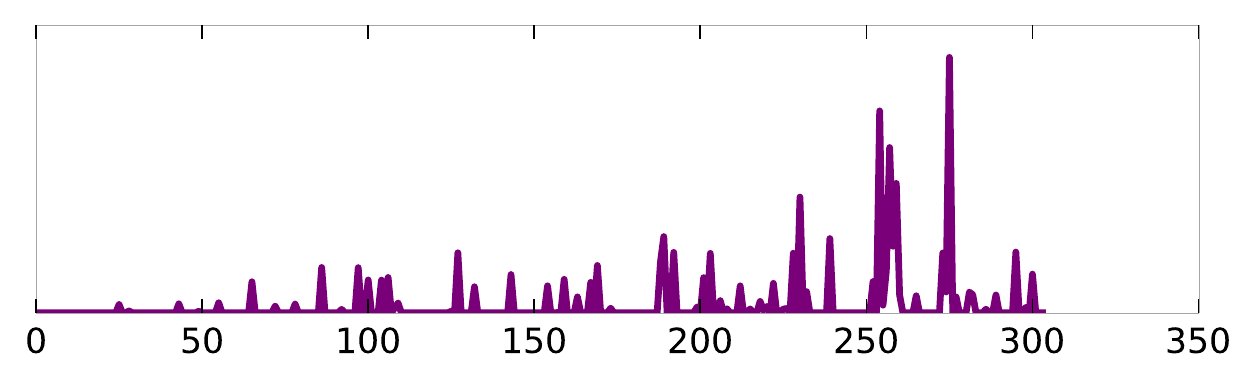}
                \caption{Time in Weeks}
                \label{fig:parafac_c1time}
        \end{subfigure}%
\begin{subfigure}[b]{0.31\textwidth}
                \includegraphics[width=\linewidth]{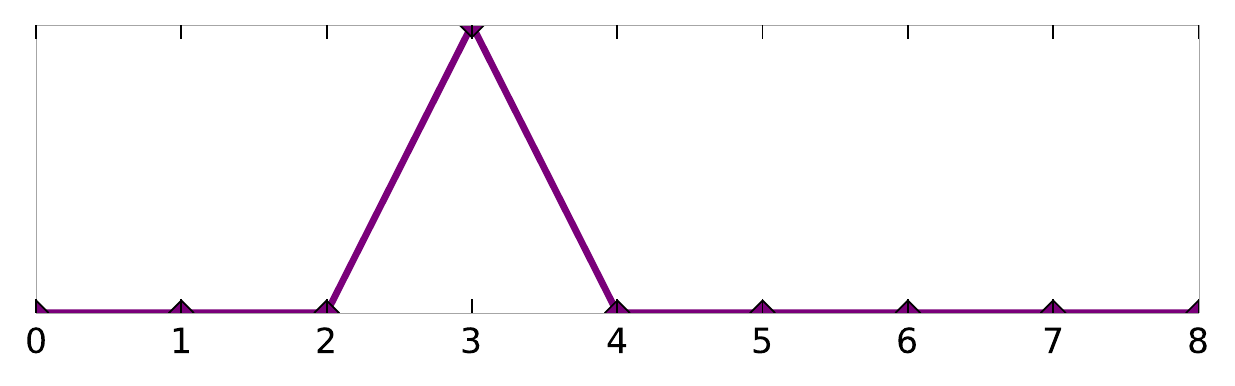}
                \caption{Log Post Number}
                \label{fig:parafac_c1post}
        \end{subfigure}%  
        
\begin{subfigure}[b]{0.31\textwidth}
                \includegraphics[width=\linewidth]{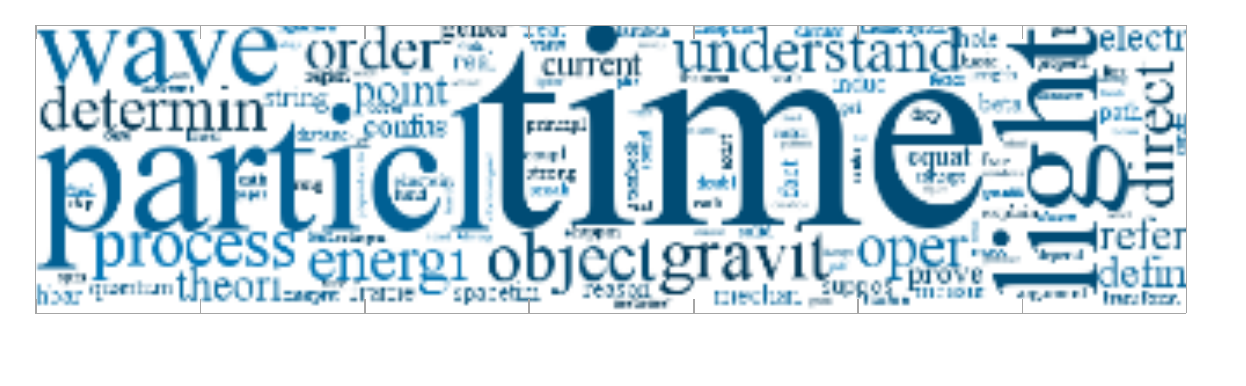}
                \caption{Words}
                \label{fig:parafac_c2word}
        \end{subfigure}%
        \begin{subfigure}[b]{0.31\textwidth}
                \includegraphics[width=\linewidth]{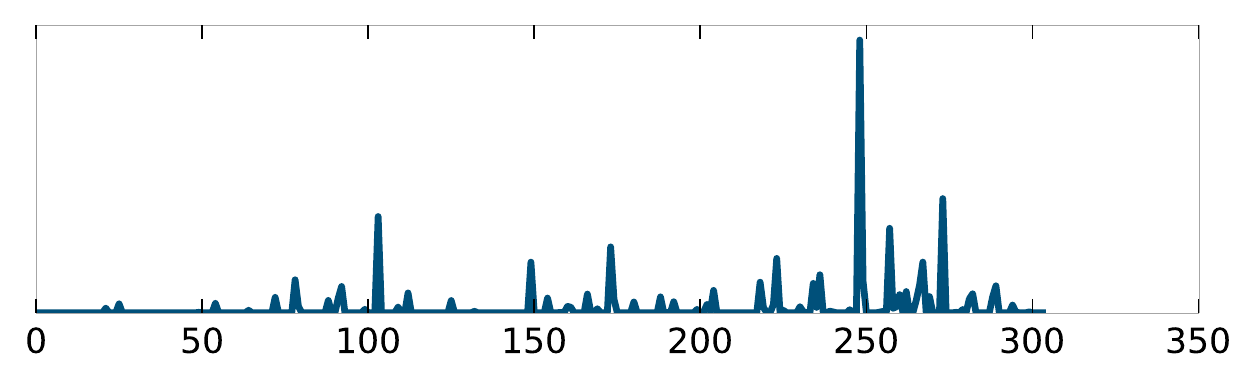}
                \caption{Time in Weeks}
                \label{fig:parafac_c2time}
        \end{subfigure}%
\begin{subfigure}[b]{0.31\textwidth}
                \includegraphics[width=\linewidth]{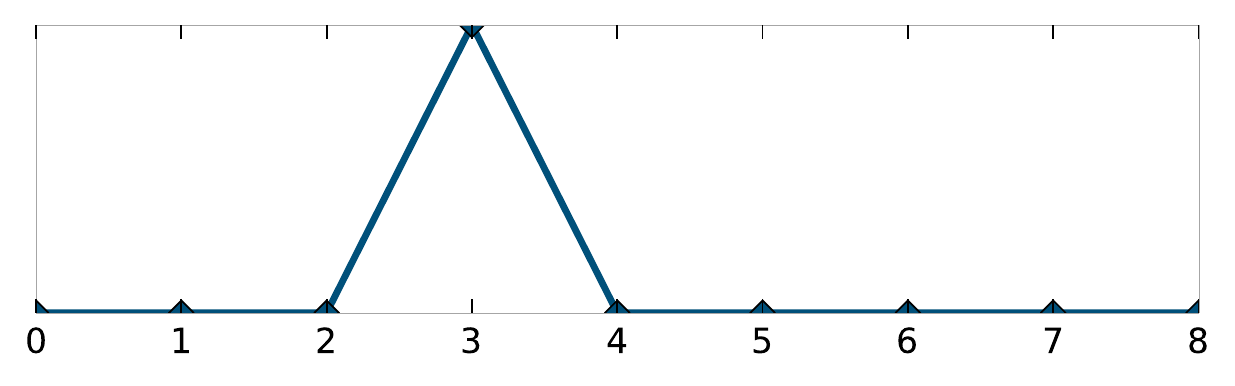}
                \caption{Log Post Number}
                \label{fig:parafac_c2post}
        \end{subfigure}%           
        \caption{\label{fig:res_parafac}An example of two components extracted by \paraNS\ algorithm on Physics dataset. The two components are similar in word, time and post number modes. The words gravity, time, light, speed, wave, particle, and energy are frequent in both components. }
\end{figure*}

\vspace{-0pt}

\begin{figure*}
\captionsetup[subfigure]{aboveskip=-1pt,belowskip=-1pt}
  \centering   
\begin{subfigure}[b]{0.31\textwidth}
                \includegraphics[width=\linewidth]{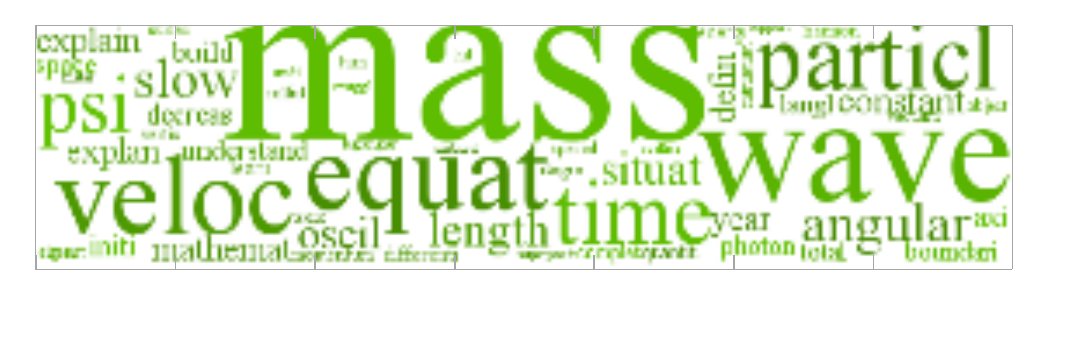}
                \caption{Words}
                \label{fig:coup_parafac_c1word}
        \end{subfigure}%
        \begin{subfigure}[b]{0.31\textwidth}
                \includegraphics[width=\linewidth]{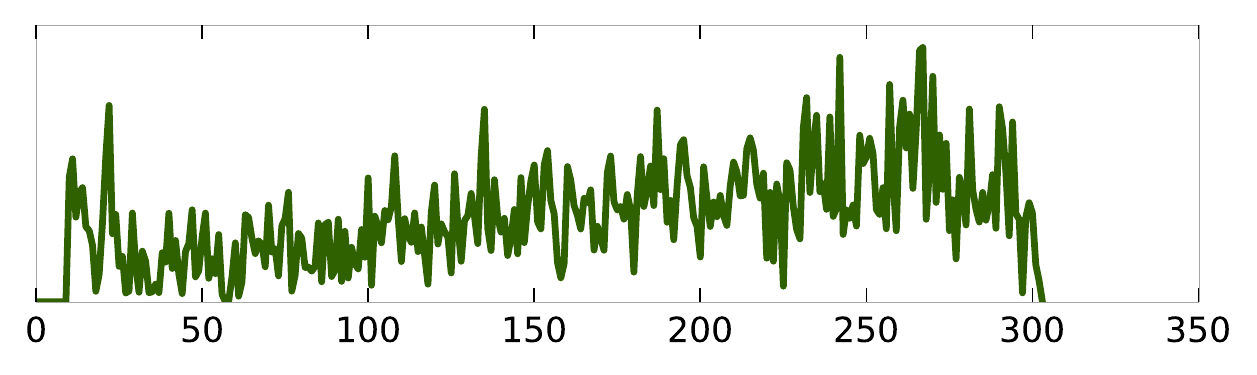}
                \caption{Time in Weeks}
                \label{fig:coup_parafac_c1time}
        \end{subfigure}%
\begin{subfigure}[b]{0.31\textwidth}
                \includegraphics[width=\linewidth]{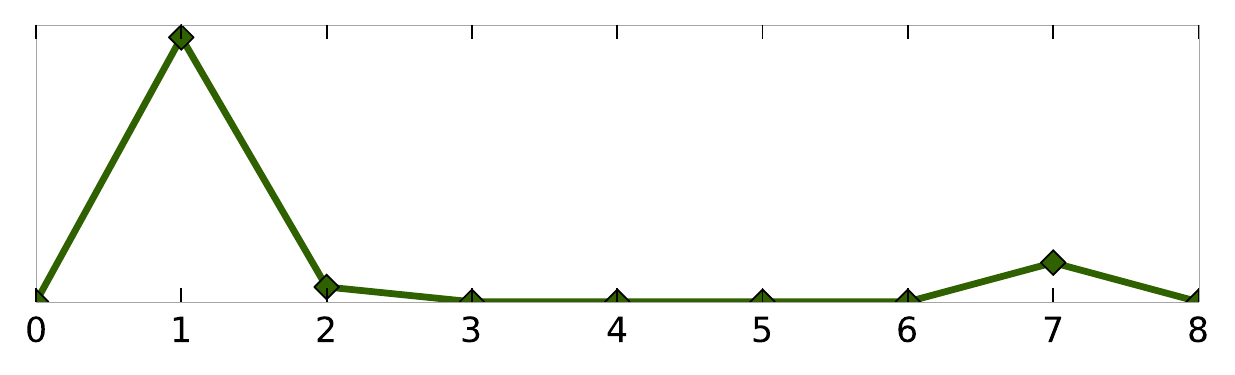}
                \caption{Log Post Number}
                \label{fig:coup_parafac_c1post}
        \end{subfigure}%   
               
                \begin{subfigure}[b]{0.31\textwidth}
                \includegraphics[width=\linewidth]{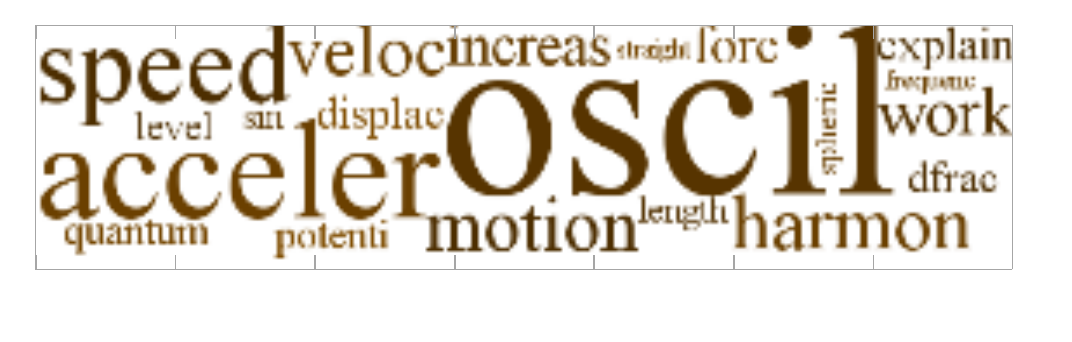}
                \caption{Words}
                \label{fig:coup_parafac_c2word}
        \end{subfigure}%
        \begin{subfigure}[b]{0.31\textwidth}
                \includegraphics[width=\linewidth]{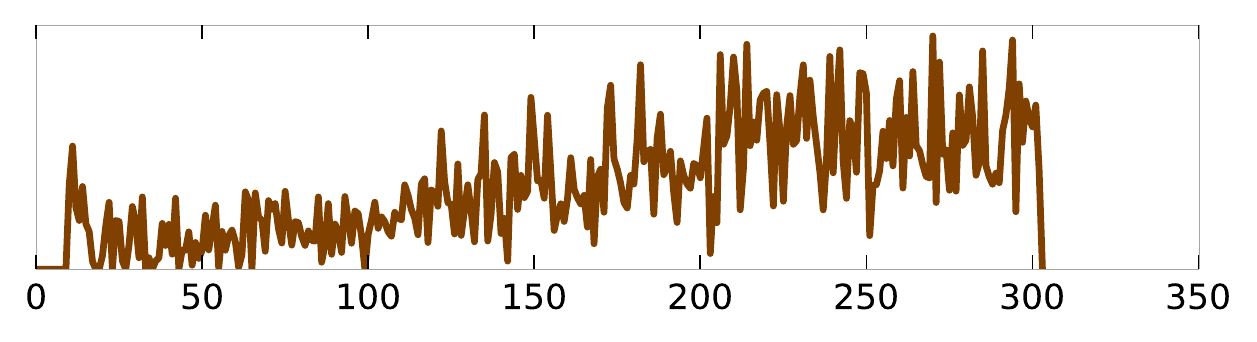}
                \caption{Time in Weeks}
                \label{fig:coup_parafac_c2time}
        \end{subfigure}%
\begin{subfigure}[b]{0.31\textwidth}
                \includegraphics[width=\linewidth]{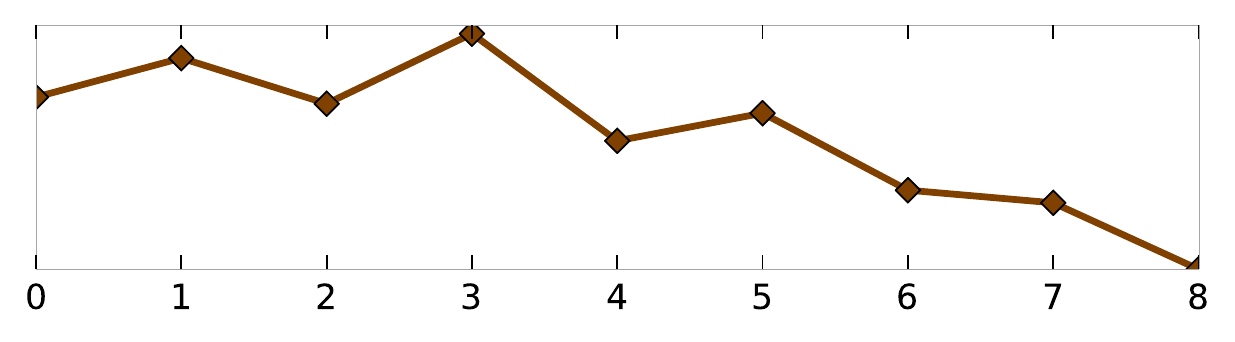}
                \caption{Log Post Number}
                \label{fig:coup_parafac_c2post}
        \end{subfigure}%
        
        \begin{subfigure}[b]{0.31\textwidth}
                \includegraphics[width=\linewidth]{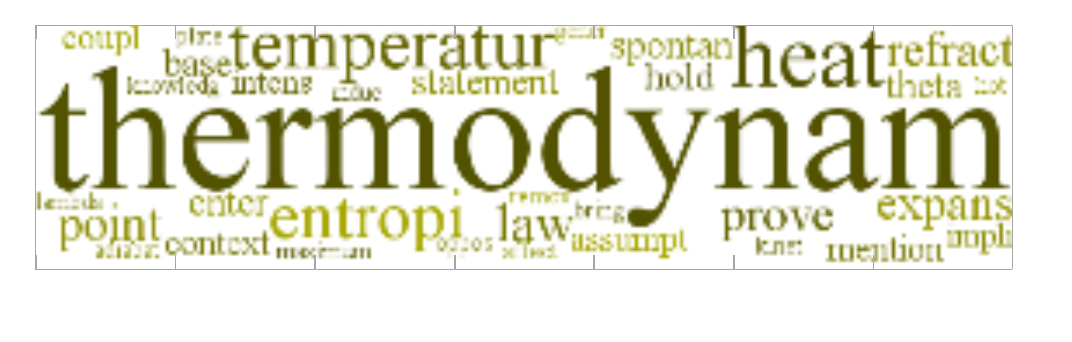}
                \caption{Words}
                \label{fig:coup_parafac_c0word}
        \end{subfigure}%
        \begin{subfigure}[b]{0.31\textwidth}
                \includegraphics[width=\linewidth]{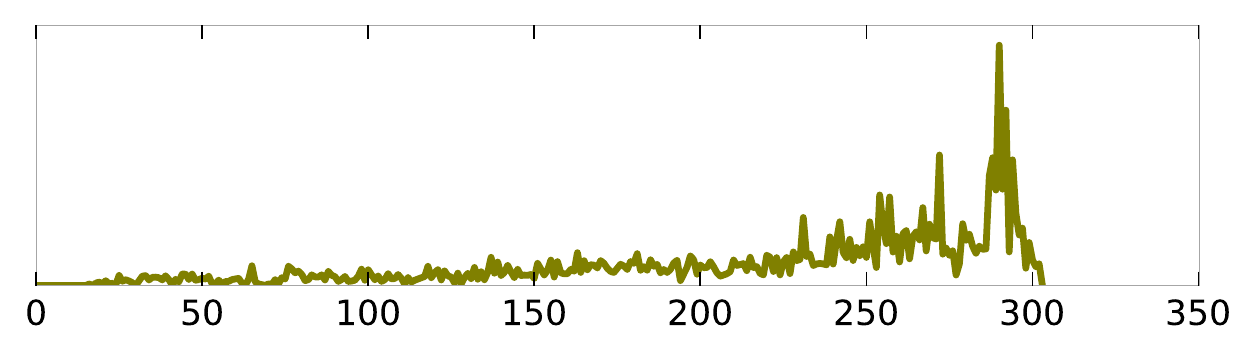}
                \caption{Time in Weeks}
                \label{fig:coup_parafac_c0time}
        \end{subfigure}%
\begin{subfigure}[b]{0.31\textwidth}
                \includegraphics[width=\linewidth]{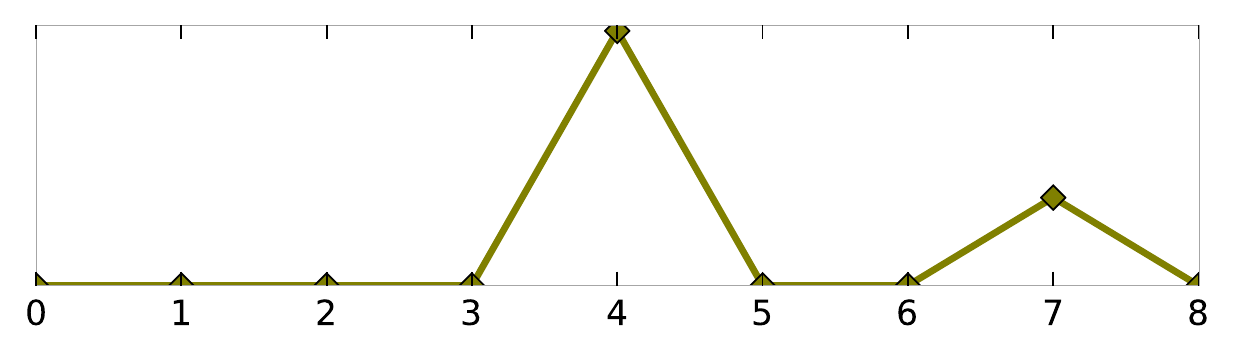}
                \caption{Log Post Number}
                \label{fig:coup_parafac_c0post}
        \end{subfigure}
        
         \begin{subfigure}[b]{0.31\textwidth}
                \includegraphics[width=\linewidth]{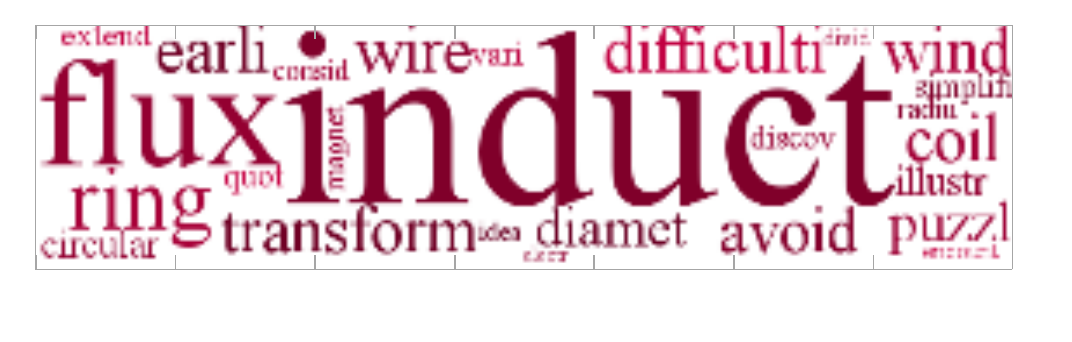}
                \caption{Words}
                \label{fig:coup_parafac_c3word}
        \end{subfigure}%
        \begin{subfigure}[b]{0.31\textwidth}
                \includegraphics[width=\linewidth]{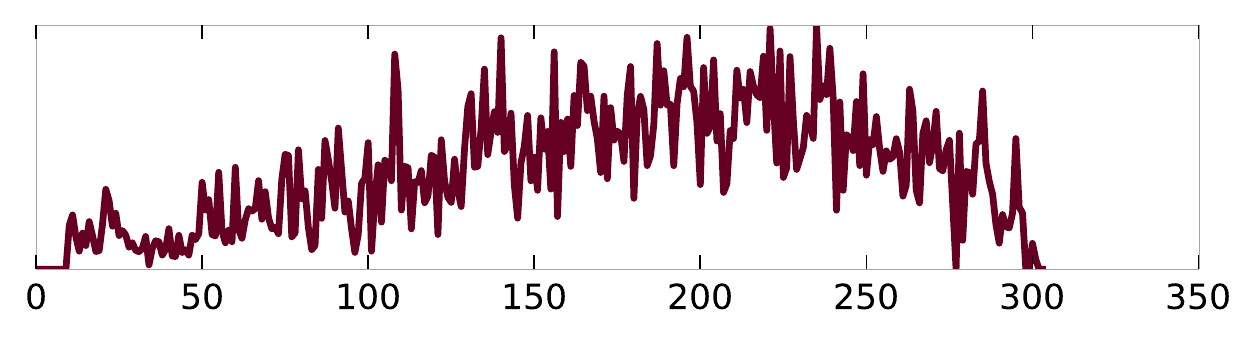}
                \caption{Time in Weeks}
                \label{fig:coup_parafac_c3time}
        \end{subfigure}%
\begin{subfigure}[b]{0.31\textwidth}
                \includegraphics[width=\linewidth]{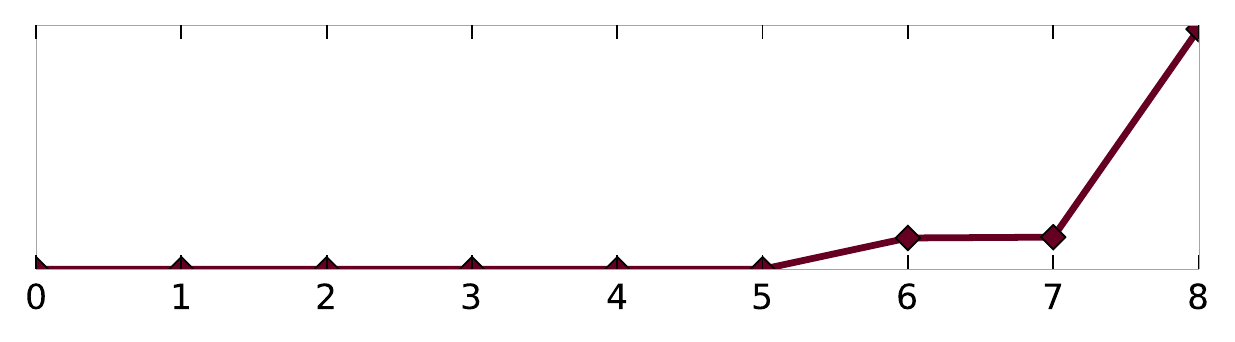}
                \caption{Log Post Number}
                \label{fig:coup_parafac_c3post}
        \end{subfigure}%
        \caption{\label{fig:coup_parafac}An example of four components extracted by \ourAlgo\ algorithm on physics dataset. All components have distinct set of words and distinct post numbers. }
\end{figure*}
\vspace{-0pt}

\begin{figure*}
\captionsetup[subfigure]{aboveskip=-1pt,belowskip=-1pt}
  \centering   
\begin{subfigure}[b]{0.31\textwidth}
                \includegraphics[width=\linewidth]{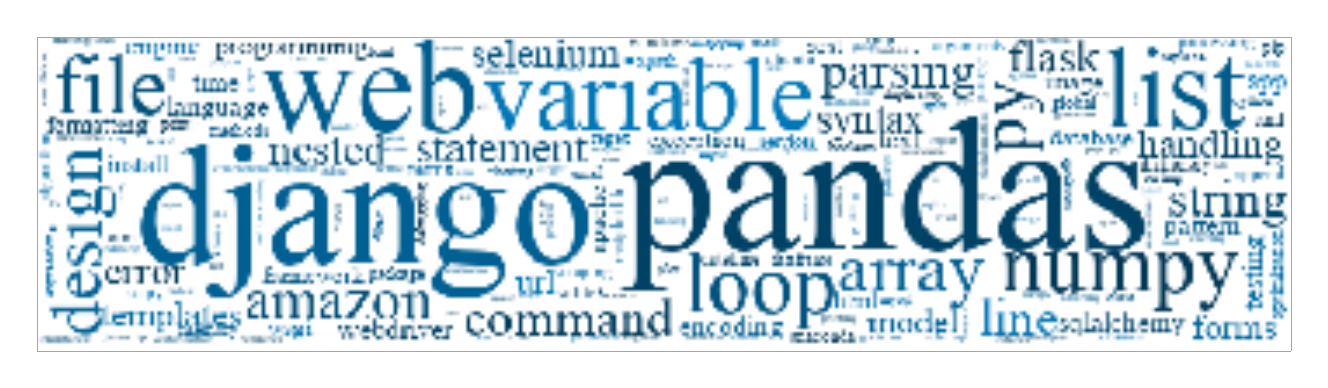}
                \caption{Words}
                \label{fig:python_parafac_c1word}
        \end{subfigure}%
        \begin{subfigure}[b]{0.31\textwidth}
                \includegraphics[width=\linewidth]{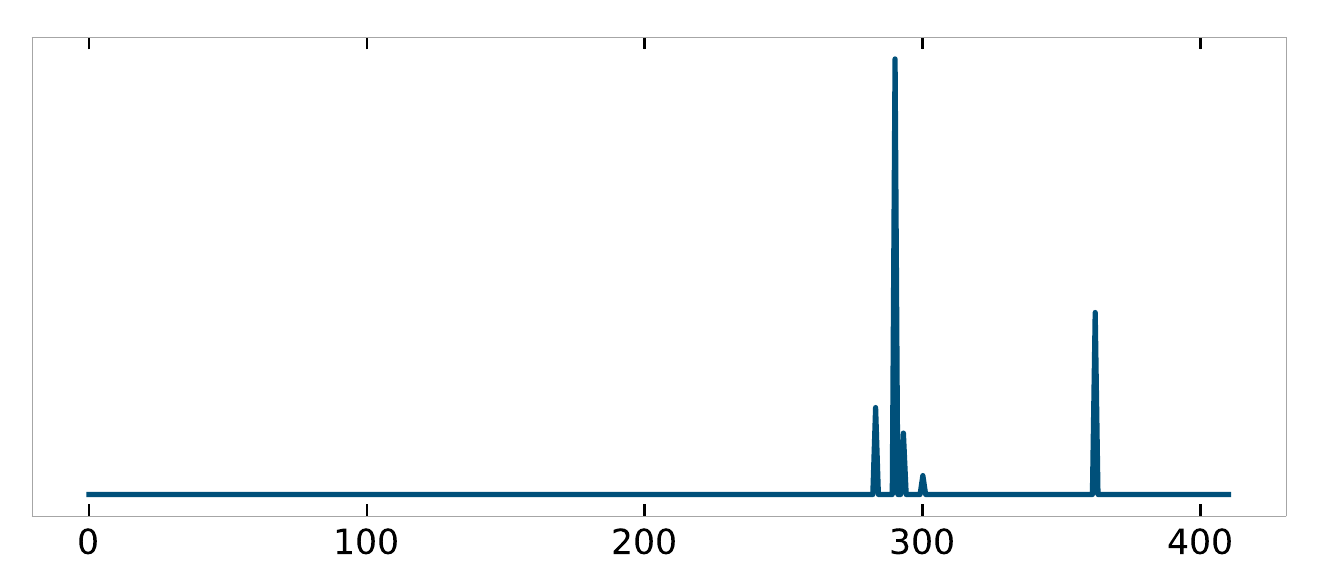}
                \caption{Time in Weeks}
                \label{fig:python_parafac_c1time}
        \end{subfigure}%
\begin{subfigure}[b]{0.31\textwidth}
                \includegraphics[width=\linewidth]{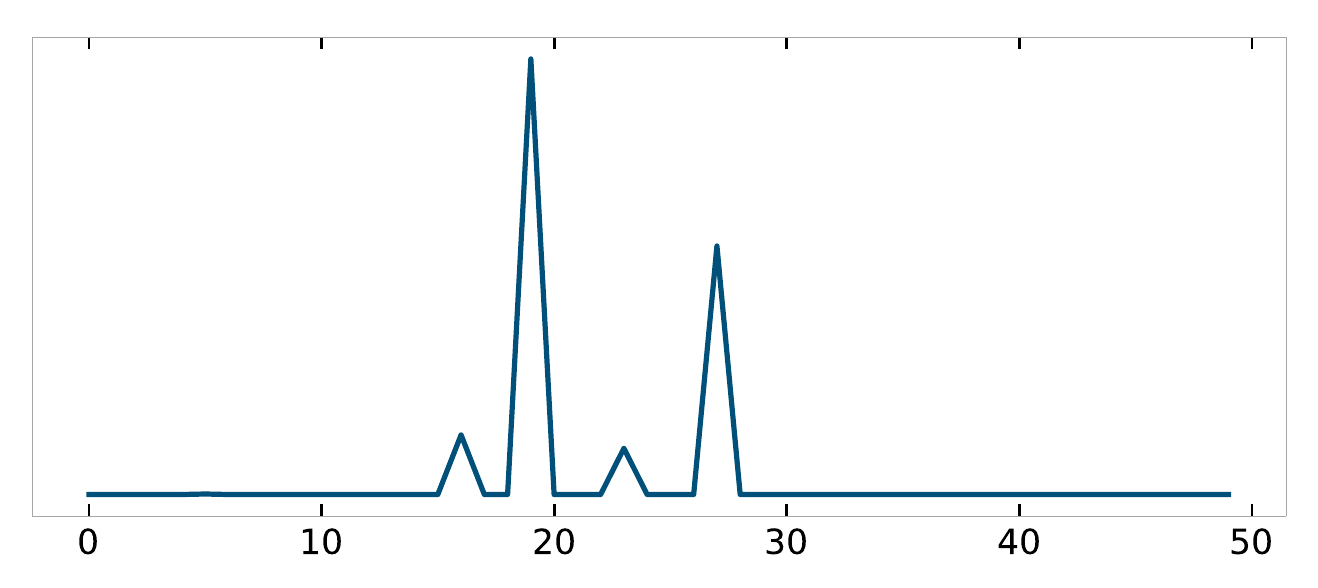} 
                \caption{Post Number}
                \label{fig:python_parafac_c1post}
        \end{subfigure}%          
        
                \begin{subfigure}[b]{0.31\textwidth}
                \includegraphics[width=\linewidth]{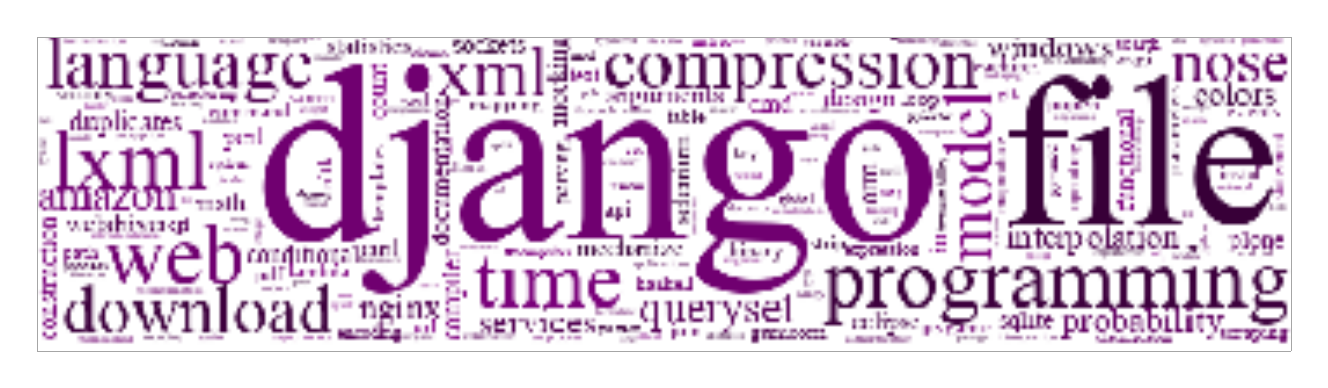}
                \caption{Words}
                \label{fig:python_parafac_c2word}
        \end{subfigure}%
        \begin{subfigure}[b]{0.31\textwidth}
                \includegraphics[width=\linewidth]{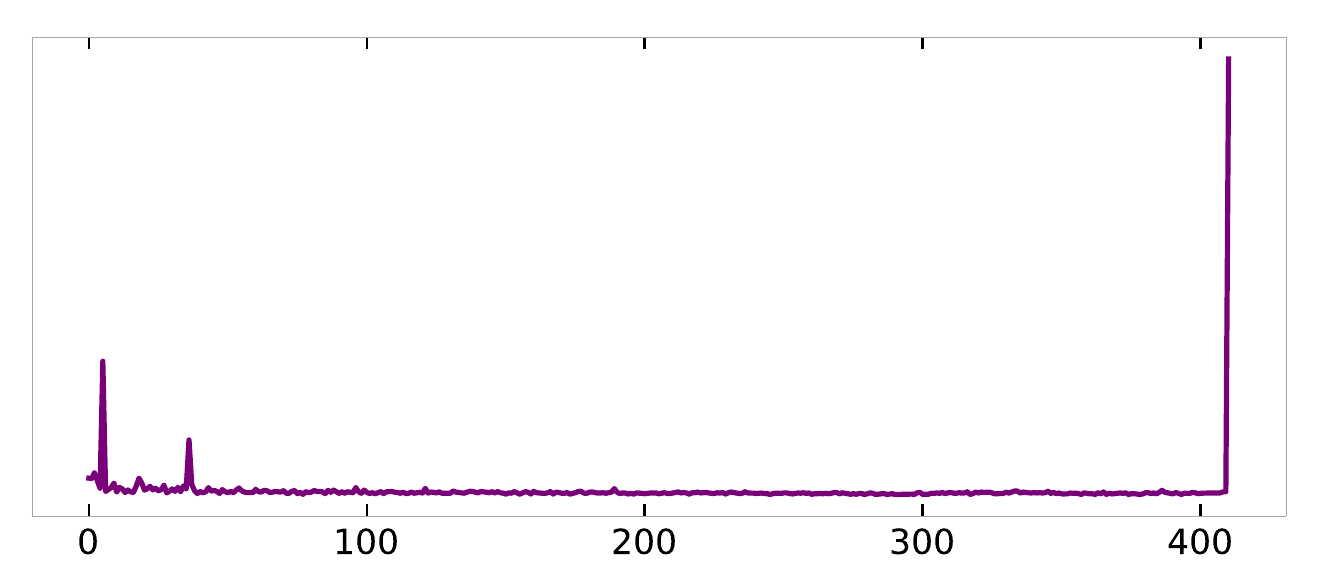}
                \caption{Time in Weeks}
                \label{fig:python_parafac_c2time}
        \end{subfigure}%
\begin{subfigure}[b]{0.31\textwidth}
                \includegraphics[width=\linewidth]{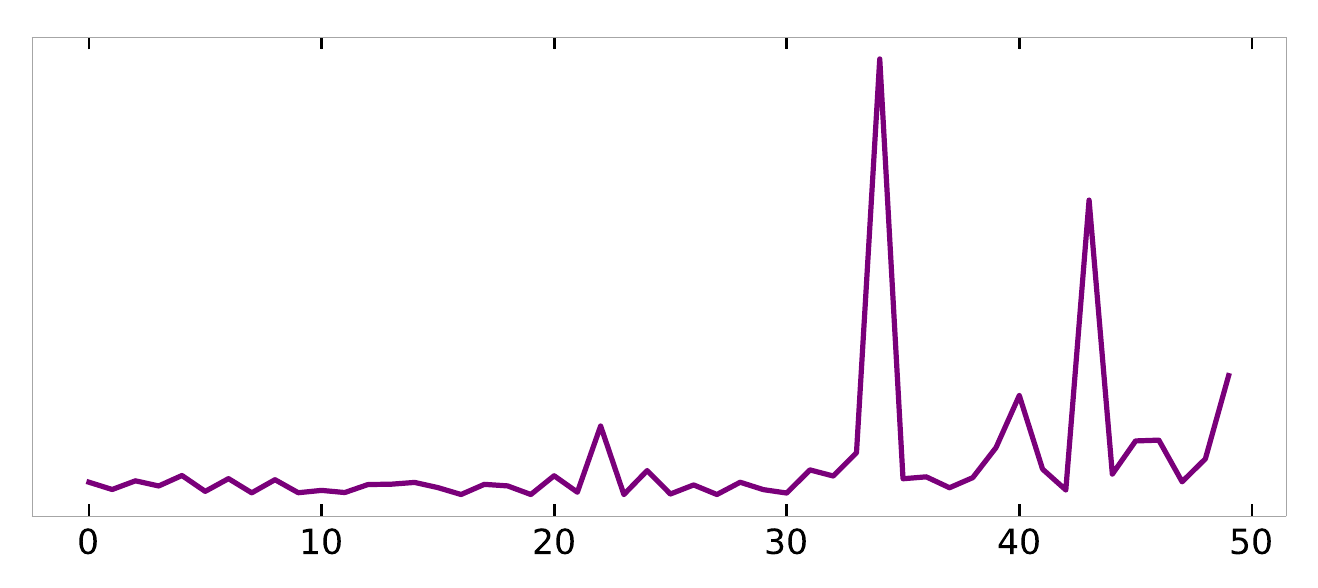}
                \caption{Post Number}
                \label{fig:python_parafac_c2post}
        \end{subfigure}%
        \caption{\label{fig:python_parafac}An example of two components extracted by \paraNS\ algorithm on programming dataset.  The two components are similar in word, time and post number modes. 
       % All components have distinct set of words and distinct post numbers. 
        }
\end{figure*}
\vspace{-0pt}

\begin{figure*}
\captionsetup[subfigure]{aboveskip=-1pt,belowskip=-1pt}
  \centering   
\begin{subfigure}[b]{0.31\textwidth} 
                \includegraphics[width=\linewidth]{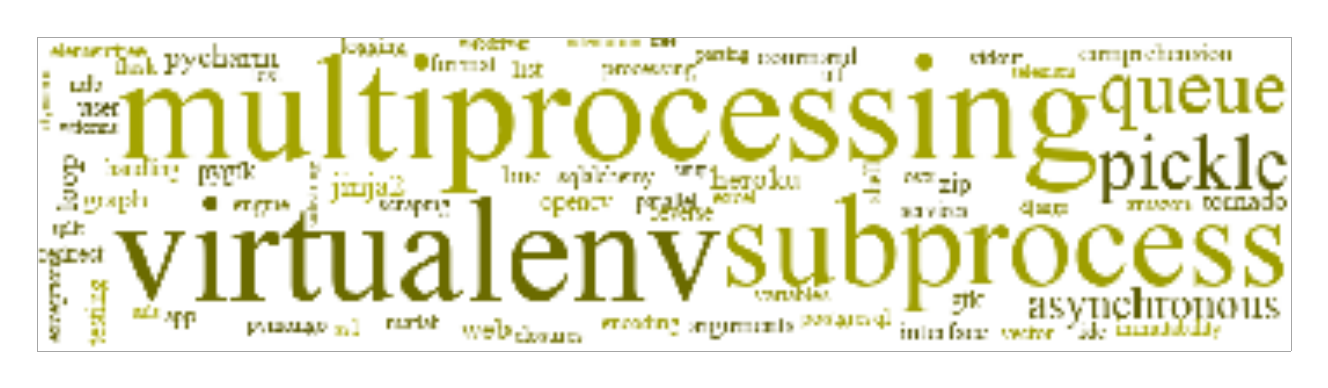}
                \caption{Words}
                \label{fig:python_coup_parafac_c1word}
        \end{subfigure}%
        \begin{subfigure}[b]{0.31\textwidth}
                \includegraphics[width=\linewidth]{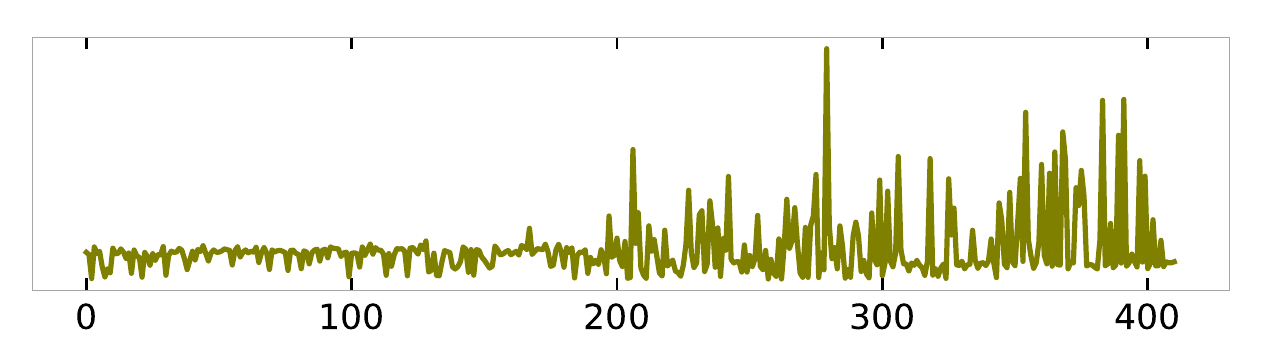}
                \caption{Time in Weeks}
                \label{fig:python_coup_parafac_c1time}
        \end{subfigure}%
\begin{subfigure}[b]{0.31\textwidth}
                \includegraphics[width=\linewidth]{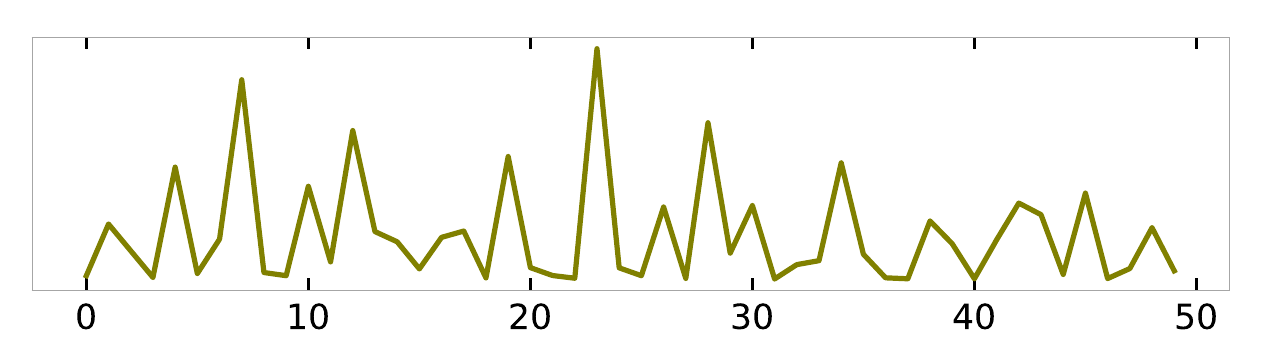} 
                \caption{Post Number}
                \label{fig:python_coup_parafac_c1post}
        \end{subfigure}%          
        
                \begin{subfigure}[b]{0.31\textwidth}
                \includegraphics[width=\linewidth]{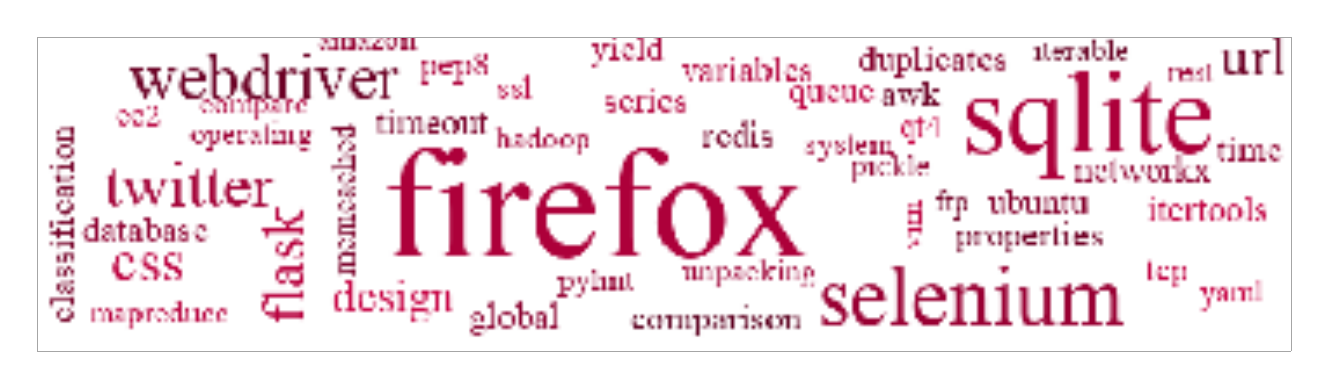}
                \caption{Words}
                \label{fig:python_coup_parafac_c2word}
        \end{subfigure}%
        \begin{subfigure}[b]{0.31\textwidth}
                \includegraphics[width=\linewidth]{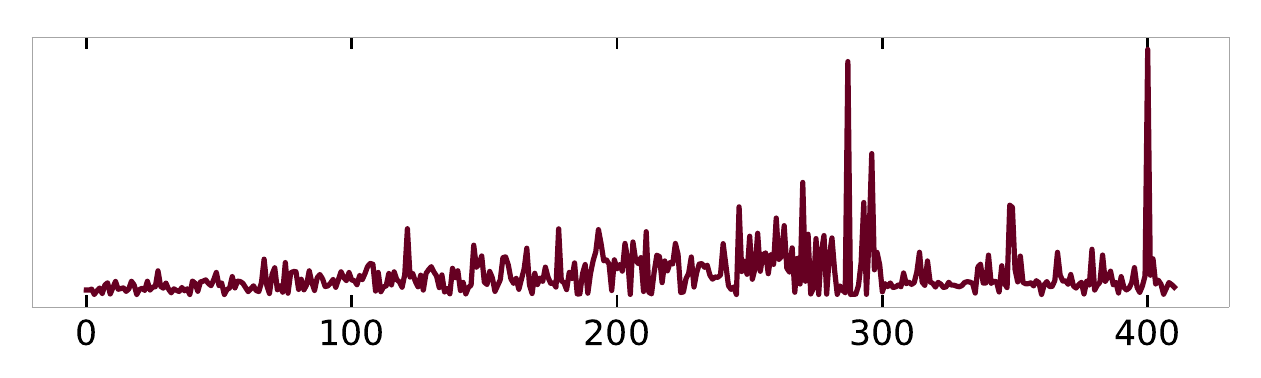}
                \caption{Time in Weeks}
                \label{fig:python_coup_parafac_c2time}
        \end{subfigure}%
\begin{subfigure}[b]{0.31\textwidth}
                \includegraphics[width=\linewidth]{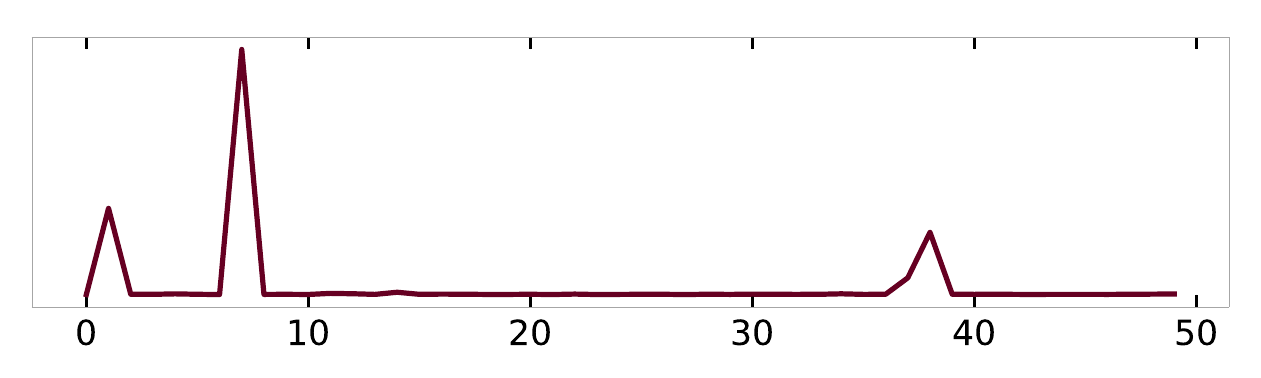}
                \caption{Post Number}
                \label{fig:python_coup_parafac_c2post}
        \end{subfigure}%
        \caption{\label{fig:python_coup_parafac}An example of four components extracted by \ourAlgo\ algorithm on programming dataset. 
       % All components have distinct set of words and distinct post numbers. 
        }
\end{figure*}
\vspace{-0pt}

\texttt{Stack Exchange} is a question answering website created in 2008. It features questions and
answers on a wide range of topics from mathematics and programming to cooking and movies.  \texttt{Stack
Exchange} allows each question  to be annotated with one or more terms (tags) indicating the subject
matter of the question.  
%We collected all 410740 questions which were tagged as a Python question from 2008 to 2016.
We used the latest Physics and programming Data Dump \
\footnote{//archive.org/details/stackexchange} in \textbf{\texttt{Stack
Exchange}}. We only consider the questions which have at least one tag (almost $30\,000$ questions),  and
we only considered frequent words which appeared more than 100 times in the forum. We also stemmed all the words. 

\subsubsection{\textbf{Physics Stack Exchange}}
 From the physics forum data, we created a tensor 
(multi-way array) \tensor\ with three modes (word, time, post number) of size $1351  \times 304 \times 9$.
When a user \user\ uses word \word\ at week \tim\ in his $p^{th}$ post, we will increase $\tensor ( \word,\tim, \log(p))$.
Thus, the $(i,j,k)$ value of Tensor \tensor\ indicates how many times word $i$ was used at week $j$ in $\log(k^{th})$ posts of
all users. Note that post number is relative to each users' sign up date. Hence, if a user
signs up and writes a question or answer, her post number would be 1.

An important aspect of considering time along with topics is the fact that the temporal information of
the topics helps us understand the topic better. For example, the word "jobs" relates to employment,
but after  October 5, 2011, the word jobs may refer to "Steve Jobs". This is the key reason for why
we use time as another dimension, we can get more insight about topics and distinguish evolutionary
topics from event driven topics which follow a bursty distribution.

In our application, beside the words, post numbers and time stamps, we also have the tags associated
to each question by the users. We can use question tags as a side information or metadata as a word-tag
matrix. This matrix indicates how many times each word has been used for a specific tag. We denote this
matrix by \sideMatrix. Our auxiliary matrix \sideMatrix, has two modes (words and tags) of size
$1351 \times 527$.  $\sideMatrix_{i,j}$  indicates the number of times word $i$ has been used in a post
with tag $j$.

\subsubsection{\textbf{Programming Stack Exchange}}
The programming Stack Exchange forum includes all the questions related to programming including Java, Python, C, R and other programming languages. We decided to only include the questions including "python" tag.  
% explain the reason
From the programming questions having python tag, we created a tensor 
(multi-way array) \tensor\ with three modes (word, time, post number) of size $432  \times 411 \times 50$.
When a user \user\ uses word \word\ at week \tim\ in his $p^{th}$ post, we will increase $\tensor ( \word,\tim, p)$. Beside the words, post numbers and time stamps, we also have the tags (except python tag) associated to each question by the users.  Similar to Physics data, we created an auxiliary word-tag matrix which indicates how many times each word has been used for a specific tag. The auxiliary matrix \sideMatrix, has two modes (words and tags) of size
$432 \times 30$ (we only kept top 30 tags).

\subsection{Experimental Evaluation}
 In this part, we evaluate our algorithms under CP/PARAFAC and Tucker3 decomposition models for CMTF.
 We implemented our algorithm in Matlab and used CVX package to solve each step of Algorithm \ref{algo_als}.
 All experiments were carried out on a machine with a 2.4 GHZ CPU, 16 GB RAM, running CentOS Linux 7. 
 \textbf{ Our dataset and our code are immediately and freely available for download}\footnote{
     \url{https://github.com/ConCMTF/ConCMTF}}
 .We compare our results to non-negative PARAFAC decomposition \cite{morup2006sparse} and sparse
 non-negative Tucker3 \cite{papalexakis2012parcube}. We refer to them  as \paraNS\ and \tuckerNS\
 respectively. To decide the right number of latent factors ($F$) to be extracted in each algorithm,
 we used AutoTen \cite{papalexakis2016automatic} which allows us to find more structured latent
 embeddings in the data. For each component that is obtained from each decomposition, we do 2-means
 clustering on the vector associated with the word mode. Then, we take the cluster with the maximum
 mean and choose the cut-off value to be equal to the smallest value in that cluster, such that any
 value below that threshold is zeroed out. In this way, we avoid interpreting the noise words (i.e.
 those with very small values) as part of the topics.

\textbf{\paraNS\ vs. \ourAlgo\ with PARAFAC:}\\
Figure \ref{fig:res_parafac} shows two components selected from obtained components  using \paraNS\
algorithm on Physics dataset. We observe that in these two decompositions, there are overlaps in the set of words found
by \paraNS\  as well as overlap in time and post number modes. In fact,  post numbers have identical
trends  and the words gravity, time, light, speed, wave, fraction, particle, and energy are among
frequent words in both components. Moreover, the set of words in both components include a (relatively)
large number of words and the word factors are very dense. If the goal of factorization is to find latent
structure and patterns in the data, these two components are very similar and hence give us the same
structure and little information.

We also used our algorithm, \ourAlgo, assuming a CP/PARAFAC decomposition.  For this decomposition, we
only imposed non-negativity and orthogonality constraint on components \factA, \factB, \factC, and
\factD\ with \epsOrthogA\ = 0.05, \epsOrthogB\ = 0.6, \epsOrthogC\ = 0.2, and \epsOrthogD\ = 0.2. The
intuition behind this is that we we would like to find components which are distinct in their set of
words and the level of maturity (post number values). However, we allow decompositions to have overlap
in the time mode as we seek patterns in any period of forums lifetime. 

Figure \ref{fig:coup_parafac} illustrates the components produced by \ourAlgo  on Physics dataset. The set of words in each component are sparse and they do not share many words  as it was in the case
in \paraNS\ components. The post numbers of each component are non-overlapping as well. The first word
component depicts the words ``mass", ``wave", ``equation", ``velocity",  ``particle", ``time", 
``angular", ``slow",  and ``oscillation"  which were used in very low post number (i.e. by
new users). These are in fact basic topics in physics. The second component  covers topics related to
harmonic motion and waves topics. Compared to the first component these words appear in larger post
numbers, i.e. they are posted by more advanced users. The topic of the third component is mainly
the first law of thermodynamics discussed mainly by advanced users with large post numbers. The last
component included the words ``inductor", ``flux", ``ring", ``diameter", ``transform", ``collide",
``magnetic", ``field", and ``circular". These words are related to "Toroidal inductors and transformers"
which only appeared in very large post number and by very advanced users.

\begin{figure}
\begin{center}
 \begin{subfigure}[b]{0.23\textwidth}
                \includegraphics[width=\linewidth]{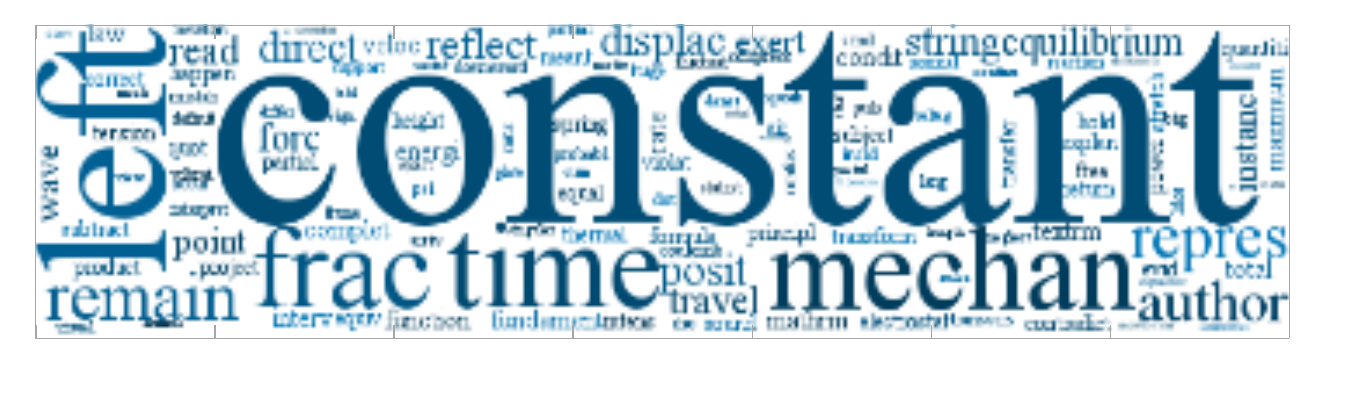}
                \caption{\tuckerNS}
                \label{fig:tucker_0word}
        \end{subfigure}%
        \begin{subfigure}[b]{0.23\textwidth}
                \includegraphics[width=\linewidth]{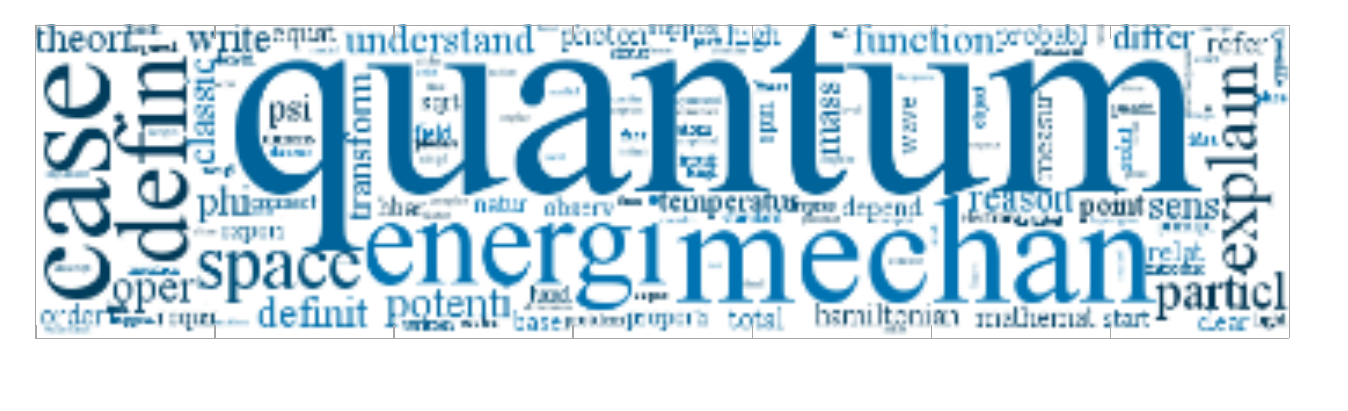}
                \caption{\tuckerNS}
                \label{fig:tucker_1word}
        \end{subfigure}%
        
\begin{subfigure}[b]{0.23\textwidth}
                \includegraphics[width=\linewidth]{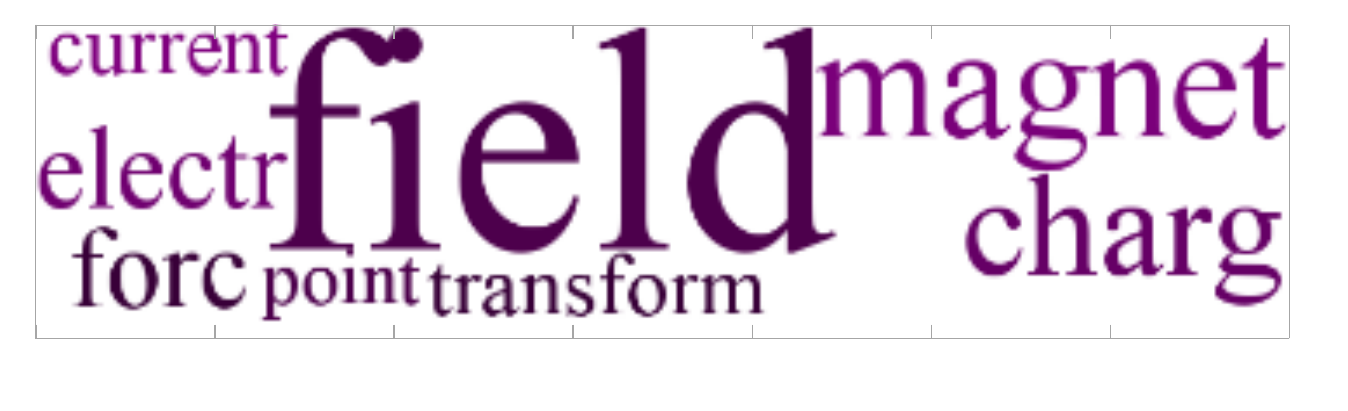}
                \caption{\ourAlgo}
                \label{fig:coup_tucker_0word}
        \end{subfigure}%  
        \begin{subfigure}[b]{0.23\textwidth}
                \includegraphics[width=\linewidth]{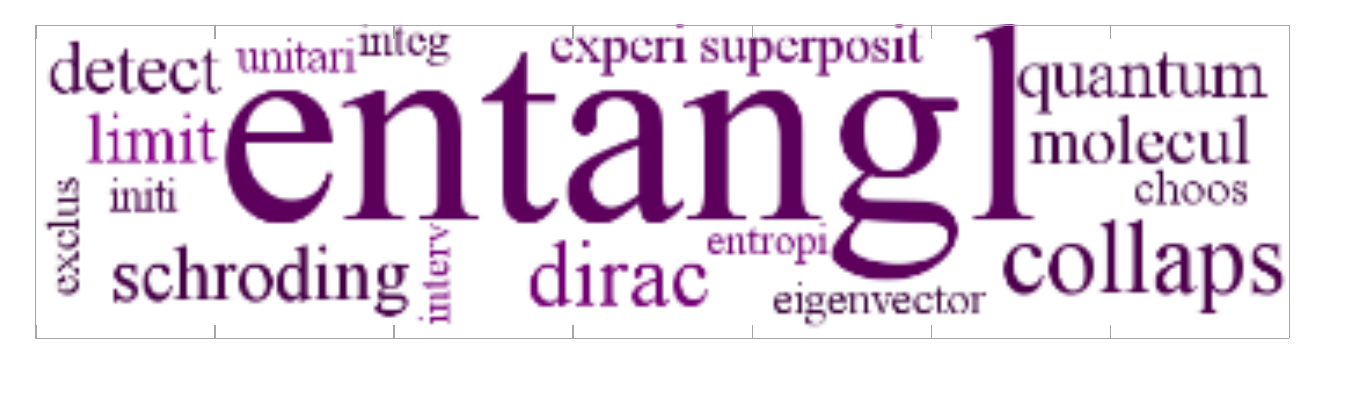}
                \caption{\ourAlgo}
                \label{fig:coup_tucker_1word}
        \end{subfigure}%  
 \caption{\label{fig:res_tucker}An example of two components extracted by \tuckerNS\ and  \ourAlgo\ algorithm. }
 \end{center}
\end{figure}

Figure \ref{fig:ligo} is an example of a component which only appeared in a specific time period and
moreover in specific post numbers (large post numbers).  This pattern indicates words discussed in
response to an external event.  The set of words in this component consists of ``gravity", ``Ligo",
``detection", ``laser", ``hole", ``theory", ``space", ``time", ``mass", etc. The peak in time mode
corresponds to Feb, 2016. This is the time that the detection of gravitational waves was
announced by Ligo lab.

Figure \ref{fig:res_parafac} shows two components selected from obtained components  using \paraNS\
algorithm on Programming dataset. As illustrated, these two decompositions, there are overlaps in the set of words found by \paraNS\  as well as overlap in time and post number modes. On the other hand, figure \ref{fig:python_coup_parafac} shows two components extracted by our algorithm.  As illustrated in the figure, 
the set of words in each component are sparse and they do not share many words  and each component shows semantically coherent topics. The first topic, figure \ref{fig:python_coup_parafac_c1word} ,includes words related to multiprocessing in python such as thread, multiprocessing subprocess, queue, virtualenv, asynchronous, and etc.  Figure \ref{fig:python_coup_parafac_c1post} shows multiprocessing topics have a presence  across various post numbers. This reveals that such a topic is of interst regardless the expertise of users.  The second topic, figure  \ref{fig:python_coup_parafac_c2word},  includes topics related to web crawling such as selenuim, web-driver, Firefox, flask, Twitter, url, css, and etc. The associated post number, figure \ref{fig:python_coup_parafac_c2post} releavs that this topic is mainly of interest of new users with lower experience. 

\textbf{\tuckerNS\ vs \ourAlgo\ with Tucker3:}\\
Figure \ref{fig:res_tucker} illustrates the decompositions obtained by \tuckerNS\  and \ourAlgo\ when assuming a Tucker decomposition with 12, 4, and 4 latent factors for each mode.
 We can observe that the  words produced  in \tuckerNS\ algorithm are not distinct and have a big overlap (Figure \ref{fig:tucker_0word} and \ref{fig:tucker_1word}). 
On the other hand, our algorithm \ourAlgo, is 
able to find more distinct and coherent set of words
 (Figure \ref{fig:coup_tucker_0word} and \ref{fig:coup_tucker_1word}. 
Moreover, using \tuckerNS, the core tensor has values in almost $55\%$ 
of the core entries, making the interpretation difficult. 
On the contrary, only $20\%$ of the core entries have valueses using \ourAlgo. 

%We sorted the words based on their frequency such that most frequent words appear with higher indices in word mode and as we see most of the latent factors have only values for frequent words. This is a natural result of  \tuckerNS\ since it tried to only optimize the error, meaning optimizing for large values in the tensor (frequent words). As a result  \tuckerNS\ in not very successful in finding the hidden structure of the data. On the other hand, our algorithm \ourAlgo, is able to find more distinct latent factors for all modes including the word mode. The results are depicted in figure. \reminder{and sparsity on the core tensor, mention the amount of sparsity}

\section{User Study}
To evaluate the quality of the topics found by our algorithm, we conducted a user study with two goals: 1) evaluate the cohesion of each learning unit, and 2) evaluate the ordering of the units. In the following sub-sections we present the details of our conducted user-study and the results of our study.

%\subsection{Survey Participants}
\textbf{Survey Participants:}
We recruited 10 volunteers to participate in our survey. Six had a PhD in computer science. Two with a PhD in Physics and Space Physics respectively, and finally two participant had mechanical engineering background.
None of the authors provided any judgements.

%\subsection{Survey Design}
\textbf{Survey Design:}
We selected 5 components from the set of all components resulted from \ourAlgo\ assuming a CP/PARAFAC decomposition.  The reason we decided to  include a limited number of components in the study is that we noticed  some set of the words associated with some components are quite  advanced and the participants are most likely not familiar with them. For example, the following component was excluded from the survey: ``bra, preserve, ket, property, configure, rigor, Hilbert, hermitian, electrodynamic". This set of words refer to  quantum mechanics and Bra-ket notation. To determine which topics are advanced and not familiar to our participants, we asked a scientist with a PhD in physics (who did not participate in our survey) to identify the less known topics.  
 We showed the remaining set of  words  to our participants and asked them to count number of odd words in each set of words produced by \ourAlgo\ assuming PARAFAC decomposition.

We asked the following question to our volunteers:

\textbf{Q1:}  Count the number of odd words in each topic.  

%\subsection{Survey Results}
\begin {table}
\footnotesize
\begin{center}
 \begin{tabular}{c c c c c c } 
 \hline
 \textbf{a} & \textbf{Min}  & \textbf{Max} & \textbf{Median} & \textbf{Mean} & \textbf{\# Concepts}\\ \hline %[0.5ex] 
 Unit 1 & 1 & 4 & 2 & 2.3 & 11  \\
 Unit 2 & 1 & 4 & 2 & 2.1 &  11\\
 Unit 3 & 0 & 1 & 0 & 0.3 &  12\\
 Unit 4 & 1 & 4 & 2 & 2.5 &  12\\
 Unit 5 & 1 & 5 & 1.5 & 2 &   13\\
 \hline
\end{tabular}
\caption {\label{table_survey}Survey results for Q1 (number of odd words in each unit)}
\end{center}
\end {table}

\textbf{Survey Results:}
Table \ref{table_survey} summarizes the results of our survey including the min, max, mean,
and the median of values that our participants reported as the number
of odd words in each topic (unit).
%Unit 1 consists of the words tidal, wide, gravity, wave, moon, force,  mass, spherical, acceleration,  newton, space, and bond. Most of these words refer to the "Moon Tides" topic in physics. 
In general, we observe that for the all of our units, the number of odd words is very low, demonstrating
good cohesion in each set of words. Unit 3 has the most number of odd words on average. This unit
consists of the words such as ``inductor, flux, ring,  difficulty, collide, avoid, wirewind, field ,
magnetic, diameter,  illustrate, circular". Most of our participants indicated ``difficulty, avoid,
illustrate, and collide" as odd words. These are the words that do not correspond to physics concepts and were not
excluded from the data. The third unit has the least number of odd words. This unit consists of words
such as mass, wave, equation, velocity, particle, time, psi, angular, slow, length, and transmit. 

It is also important to evaluate the inter-judge  agreement in a survey like ours. Due to the nature of
the ratings, an appropriate way of analyzing the agreement is by using Krippendorff’s $\alpha$
statistical measure \cite{krippendorff1970estimating}, which is applicable to the current scenario of
judges assigning a value to a specific variable. The overall agreement measured by Krippendorff’s
$\alpha$ for our ten judges turns out to be 0.32. This indicates that there is fair but imperfect
agreement. When we look at the agreement of judges within the same background group, we observe the
agreement between the second group with Physics and space physics background is 0.56 which shows our
participants with physics background have a high level agreement on the number of odd topics.

\subsection{Applicability to Curriculum Design}
As we mention in the Introduction, our proposed topic discovery has implications to curriculumn design,
since it is able to identify topics along with their level of difficulty; those levels of difficulty
are key in determining prerequisite and co-requisite (i.e., concepts that must be taught at the same
time) relations between concepts in the syllabus.
%Since our model finds topics and their relevant difficulty from existing online discussions, we can use it for generating curriculum from online discussions. 
Here, we demonstrate this applicability of our topic discovery to automated curriculum
design, along the lines of recently proposed work of \cite{agrawal2016toward, agrawal2015datadriven}. In order to achieve
this, first, we identify events triggered by external events which are not part of topics evolution
and exclude them from the curriculum. We then order the topics based on their relevant difficulty,
as inferred by our topic discovery, and as we mentioned previously, this ordering determines the
arrangement of the topics in the curriculum. What follows is a curriculum we obtained from online
discussion after removing all non-physics terms. 
\begin{mdframed}[backgroundcolor=blue!4] 
\centering
\footnotesize
Flow, Mass, Work, Density, Motion, Speed, Velocity, Displacement, Acceleration, Momentum, Gravity,
    Force, Waves, Electromagnetic, Radioactivity,  Quantum, Particles
\end{mdframed} 
This curriculum is consistent with majority of curricula taught in basic physics courses in online/traditional classrooms. See { \footnotesize \url{https://en.wikipedia.org/wiki/List_of_physics_concepts_in_primary_and_secondary_education_curricula}}. 
%We reserve further investigation of automated curriculum design based on extracted topics using our method for future work.

\section{Conclusion}   \label{conclusion}
In this paper, we proposed a novel time-evolving topic discovery method which is able to identify the level of difficulty of the extracted topics. Our approach is powered by a novel Constrained Coupled Matrix-Tensor Factorization for which we provide our code at \url{https://github.com/ConCMTF/ConCMTF}. We evaluate our resulting topics both qualitatively and quantitatively via a user study of expert judges, and we demonstrate the effectiveness of the proposed method in discovering high-quality, interpretable topics, their temporal evolution, and their level of difficulty.  Finally, we highlight the implications of our approach in education-related applications.% and for future work we intend to further investigate this direction. 

\hide{
model for automatically synthesizing curriculum without human intervention given the online discussions.   
In our application, each curriculum unit corresponds to a topic extracted from the discussion.
Our experiments on large-scale data from Stack Overflow indicate our model is capable of generating desirable curriculum both in terms of extracting important curriculum units and ordering these units.
 One can improve the proposed method by considering n-grams or only  tags of the discussions. More in-depth analysis, especially evaluating the achieved curriculum by human will be necessary to extract meaningful insights.
}

%\section*{Acknowledgment}

%\section*{References}
\bibliographystyle{abbrv}
 \balance
{\footnotesize\bibliography{main}}

\end{document}